\documentclass[prd,twocolumn,preprintnumbers,nofootinbib,superscriptaddress]{revtex4-1}
\usepackage[a4paper, hdivide={1.91cm,,1.165cm}, vdivide={1.83cm,,2.4cm}]{geometry}
\usepackage[utf8]{inputenc}

\usepackage{amsmath,amssymb}
\usepackage{graphicx,multirow}
\usepackage{color}
\usepackage{slashed}
\usepackage{units}
\usepackage[hyperfootnotes=false]{hyperref}
\usepackage{siunitx}
\usepackage{upgreek}
\usepackage{textgreek}

\def \L {\mathcal{L}} 

\def \vec#1{{\boldsymbol{#1}}}
\newcommand{\hc}{\ensuremath{\text{h.c.}}}

\newcommand{\dd}{\mathrm{d}}
\newcommand{\katrin}{\textsc{Katrin}}
\DeclareSIUnit{\cps}{cps}

\begin{document}

\title{Tritium beta decay with additional emission of new light bosons}

\preprint{ULB-TH/18-12, UCI-TR-2018-14}

\author{Giorgio \surname{Arcadi}
\email{giorgio.arcadi@mpi-hd.mpg.de}}
\affiliation{Max-Planck-Institut f\"ur Kernphysik, Saupfercheckweg 1, 69117 Heidelberg, Germany}

\author{Julian \surname{Heeck}
\email{julian.heeck@uci.edu}}
\affiliation{Service de Physique Th\'eorique, Universit\'e Libre de Bruxelles, Boulevard du Triomphe, CP225, 1050 Brussels, Belgium}
\affiliation{Department of Physics and Astronomy, University of California, Irvine, CA 92697-4575, USA}

\author{Florian \surname{Heizmann}
\email[Electronic address: ]{florian.heizmann@kit.edu}}
\affiliation{Institute of Experimental Particle Physics, Karlsruhe Institute of Technology, Wolfgang-Gaede-Str.~1, 76131~Karlsruhe, Germany}

\author{Susanne \surname{Mertens}
\email[Electronic address: ]{mertens@mpp.mpg.de}}
\affiliation{Max Planck Institute for Physics, F\"{o}hringer Ring 6, 80805 M\"unchen, Germany}
\affiliation{Technische Universit\"at M\"unchen, Arcisstraße 21, 80333 M\"unchen, Germany}

\author{Farinaldo S.\ \surname{Queiroz}
\email{farinaldo.queiroz@mpi-hd.mpg.de}}
\affiliation{International Institute of Physics, Federal University of Rio Grande do Norte, Campus Universit\'ario, Lagoa Nova, Natal-RN 59078-970, Brazil}

\author{Werner \surname{Rodejohann}
\email[Electronic address: ]{werner.rodejohann@mpi-hd.mpg.de}}
\affiliation{Max-Planck-Institut f\"ur Kernphysik, Saupfercheckweg 1, 69117 Heidelberg, Germany}

\author{Martin \surname{Slez\'{a}k}
\email[Electronic address: ]{slezak@mpp.mpg.de}}
\affiliation{Max Planck Institute for Physics, F\"{o}hringer Ring 6, 80805 M\"unchen, Germany}

\author{Kathrin \surname{Valerius}
\email[Electronic address: ]{kathrin.valerius@kit.edu}}
\affiliation{Institute for Nuclear Physics, Karlsruhe Institute of Technology, P.O.\ Box 3640, 76021 Karlsruhe, Germany}

\hypersetup{
    pdftitle={Tritium beta decay with additional emission of new light bosons},
    pdfauthor={}
}


\begin{abstract}
\noindent 
We consider tritium beta decay with additional emission of light pseudoscalar or vector bosons coupling to 
electrons or neutrinos. The electron energy spectrum for all cases is evaluated and shown to be well estimated by approximated analytical expressions. We give the statistical sensitivity of \katrin~to the mass and coupling of the new bosons, both in the standard setup of the experiment as well as for future modifications in which the full energy spectrum of tritium decay is accessible. 

\end{abstract}

\maketitle


\section{\label{sec:intro}Introduction}
\noindent
It is usually expected that new physics effects may arise at high energies, but with more stringent collider limits on heavy new physics, focus in particle-physics phenomenology is shifting towards ``light'' 
new physics. In this paper we deal with possible new physics below 18.6\,keV, 
which is the endpoint of tritium beta decay. This decay is at the focus of direct neutrino mass experiments~\cite{Drexlin:2013lha}.  With the Karlsruhe Tritium Neutrino experiment (\katrin) \cite{Osipowicz:2001sq,Angrik:2005ep}, which is poised to start neutrino-mass measurements in 2019, the sensitivity on $m_\upnu$ will be improved by one order of magnitude to \SI{0.2}{\electronvolt} (\SI{90}{\percent} C.L.) by 
high-resolution \textbeta-decay spectroscopy at the kinematic endpoint. In addition, a wide variety of new physics can be addressed \cite{Riis:2010zm,Formaggio:2011jg,SejersenRiis:2011sj,Esmaili:2012vg,Diaz:2014hca,Rodejohann:2014eka,Steinbrink:2017uhw}. Interestingly, further plans exist to modify the experiment in order 
to access the full energy spectrum \cite{Mertens:2014nha,Mertens:2018}. While the main motivation 
was originally to probe keV-scale sterile neutrinos with possible connections 
to warm dark matter \cite{Adhikari:2016bei}, various new physics 
opportunities can actually be explored with this apparatus as well~\cite{Shrock:1980vy,Herczeg:2001vk,Severijns:2006dr,Liao:2010yx,deVega:2011xh,Abdurashitov:2014vqa,Barry:2014ika,Rodejohann:2014eka,Gonzalez-Alonso:2018omy,Abada:2018qok}.

In this work, we entertain the emission of light pseudoscalars and vector bosons off the neutrino or electron in tritium beta decay for the standard and the modified full-spectrum setup of \katrin. For all scenarios considered we will compute the exact electron energy spectra and provide approximated analytical estimates which are useful for sensitivity studies. We will determine, for both \katrin~setups, the statistical sensitivity for a set of simplified particle physics scenarios. Furthermore, we compare our findings with cosmological and astrophysical constraints to light new particle states as well as with other laboratory searches.  


The paper is organized as follows: in Sec.~\ref{sec:main} we discuss the various frameworks in which our light particles couple to electrons and neutrinos, present general considerations on the electron spectra in those cases, and give their simple analytical expressions. Section~\ref{sec:eV} comprises the analysis in the standard \katrin~setup, while the sensitivity for a modified full spectrum setup is treated in Sec.~\ref{sec:keV}. 
Other bounds on the models under study are discussed in Sec.~\ref{sec:other}, before we conclude in Sec.~\ref{sec:concl}. 
Technical details on the derivation of the four-body spectra are delegated to an appendix. 

\section{\label{sec:main}The theoretical spectrum}
\noindent The standard beta decay of tritium $^3{\rm H}$ into helium $^3{\rm He}^{+}$, an electron and an anti-neutrino, is mediated by a virtual $W$ boson:
\begin{align}
 {}^3\text{H} \to  {}^3\text{He}^{+}+ e^-+ \bar \upnu_e \,.
\end{align}
This decay gives rise to a continuous electron energy spectrum with endpoint $(m_{ {}^3\text{H}}^2-(m_{ {}^3\text{He}^{+}}+m_\upnu)^2+m_e^2)/(2 m_{ {}^3\text{H} })$, corresponding to a maximal kinetic electron energy of \unit[18.6]{keV}.
An exact analytical expression for the electron energy spectrum can 
be found in Ref.~\cite{Ludl:2016ane} (see also \cite{Masood:2007rc}). 

New particles with mass below \unit[18.6]{keV} can potentially be emitted in the decay and modify the electron energy spectrum. 
In this work we study the electron spectrum resulting from a four-body decay
\begin{equation}
{}^3\text{H} \to  {}^3\text{He}^{+} + e^-+ \bar \upnu_e + X\,,
\end{equation}
with $X$ being a new light boson, either scalar or vector in the following. 
The electron endpoint energy of this decay in all cases is 
\begin{align}
E_e^\text{max} = \frac{m_{{}^3\text{H}}^2-(m_{{}^3\text{He}^{+}}+m_\upnu+m_X)^2+m_e^2}{2 m_{{}^3\text{H}}} 
\label{eq:endpoint_energy}
\end{align}
and thus only eV-scale bosons can be addressed with the standard \katrin~setup exploring the endpoint region of tritium. 
A proposed extension of \katrin~could, however, explore the entire electron spectrum and thus be sensitive even to keV-scale bosons. 

For simplicity, we will set the neutrino mass $m_\upnu$ to zero in the remaining discussion of this section. As it turns out, moderate constraints on the coupling of the light bosons are enough to not impact the \katrin~neutrino mass sensitivity of \SI{200}{\milli\electronvolt}.

\subsection{\label{sec:Pnu}Pseudoscalars emitted from neutrinos}

\begin{figure}[t]
	\includegraphics[width=0.45\textwidth]{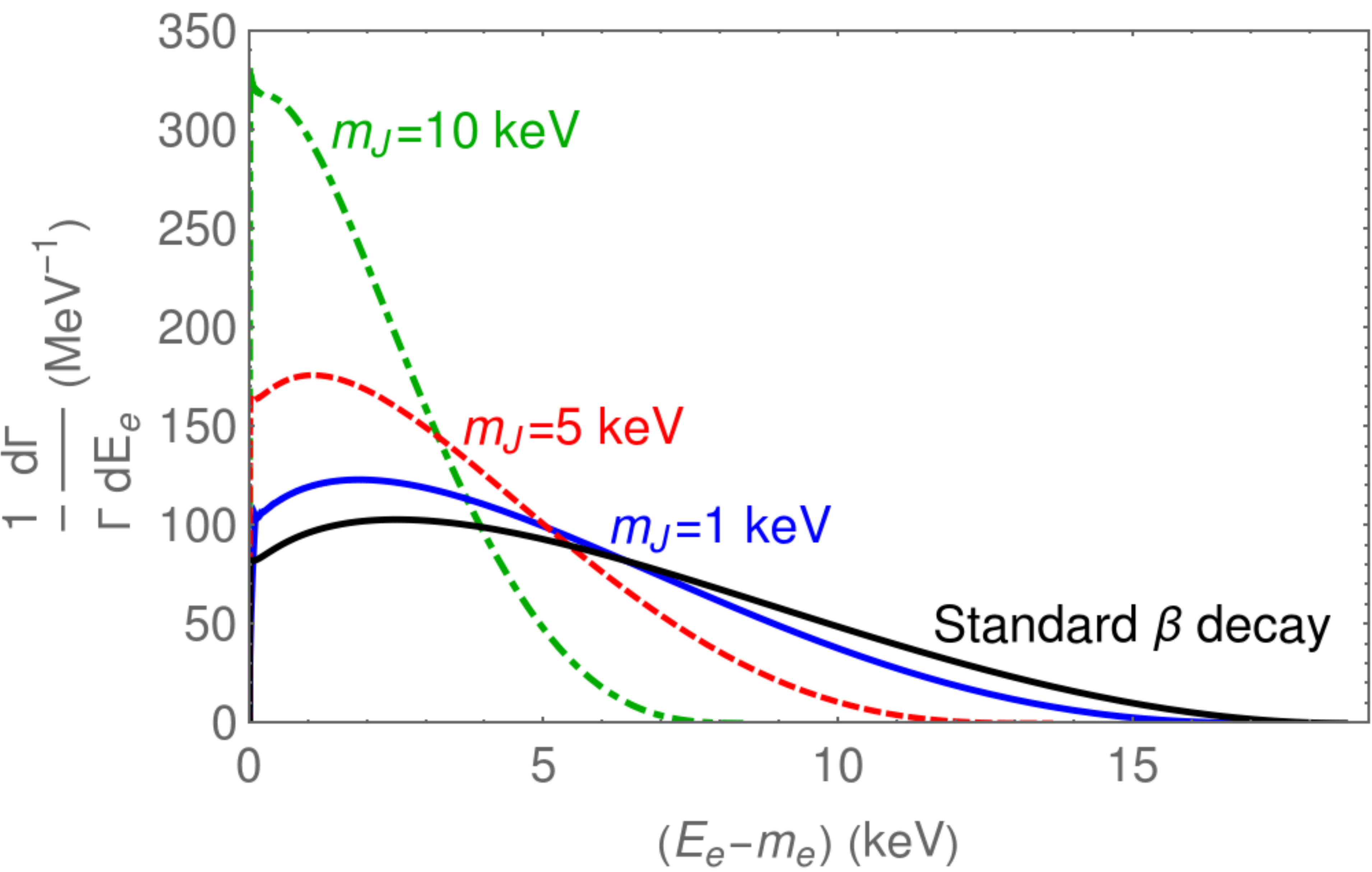}
	\caption{Normalized electron energy spectrum of the decay 
		${}^3\text{H} \to {}^3\text{He}^{+} + e^-+ \bar \nu_e + J$ (with the pseudoscalar $J$ emitted from the neutrino)  and 
		the standard $\upbeta$-decay spectrum. The coupling $g_{\nu J}$ 
		has been set to $1$, the spectrum scales with $g_{ \nu J}^2$.}
	\label{fig:vvJ-norm_spec}
\end{figure}

Pseudoscalar particles arise in many extensions beyond the Standard Model (SM), especially in those with a complex scalar sector. It is common to find massive CP-even and CP-odd (pseudoscalar) scalar fields in these theories. Axions and axion-like particles also fall into this category. In some cases, these pseudoscalar particles play the role of mediators between SM particles and the dark matter sector. In some other cases, they are connected to neutrino masses and lepton number violation, most notably in so-called Majoron models~\cite{Chikashige:1980ui,Schechter:1981cv,Pilaftsis:1993af,Garcia-Cely:2017oco}. Agnostic to its possible origin, we assume here that the pseudoscalar $J$ is coupled to neutrinos via 
\begin{equation} \label{eq:vvJ}
{\cal L} = i g_{\nu J}\, \bar \nu \, \gamma_5 \, \nu \, J \,. 
\end{equation}
The above coupling is lepton number conserving. As indicated above, there is another possibility, namely a 
lepton number violating coupling $\bar\nu^c \gamma_5 \nu J$. For $\upbeta$-decay there is no difference, while some limits depend on the nature of the coupling, see Sec~\ref{sec:other}. 
The electron energy spectrum from the decay ${}^3\text{H} \to {}^3\text{He}^{+} + e^-+ \bar \nu_e + J$ is derived in detail in App.~\ref{sec:appendix} and illustrated in Fig.~\ref{fig:vvJ-norm_spec} for various masses. Since analytical expressions are highly cumbersome, we will resort to an excellent approximate expression, given by
\begin{align}
\frac{\dd \Gamma}{\dd E_e} = \frac{K}{\hbar} \sqrt{\frac{E_e}{m_e}-1} \, \left(1- \frac{E_e}{E_e^\text{max}}\right)^n F(Z,E_e)\,.
\label{eq:approx_spectrum}
\end{align}
Here, $F(Z,E_e)$ is the Fermi function, Eq.~\eqref{eq:fermi}, which describes the interaction of the outgoing electron with the helium nucleus; $K$ is a dimensionless normalization factor and $n$ the spectral index, which is between $2$ and $4.5$ for all our models.

\begin{figure*}[t]
    \includegraphics[width=0.49\textwidth]{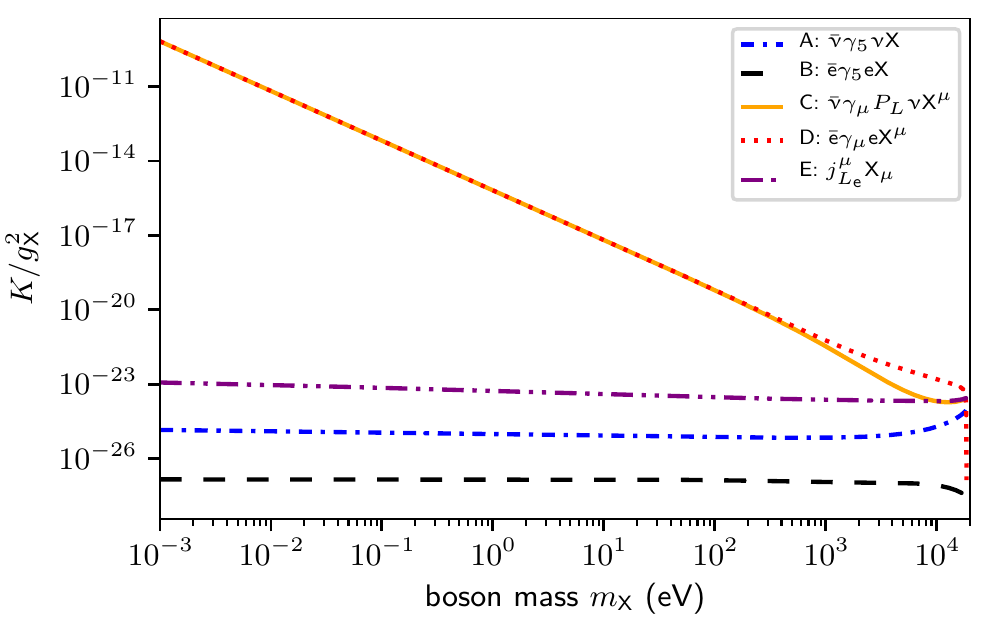}
    \includegraphics[width=0.49\textwidth]{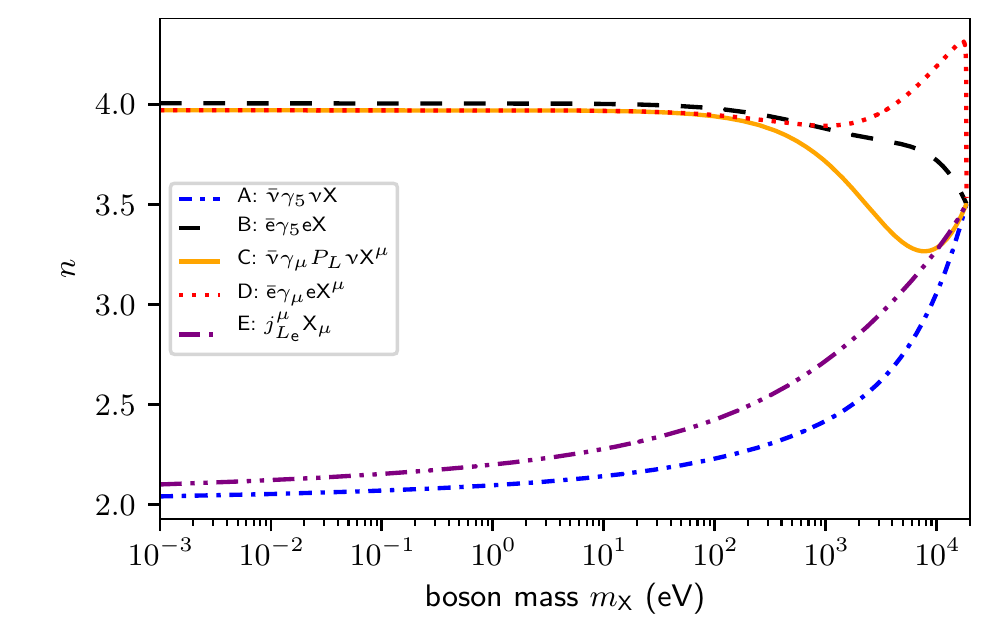}
	\caption{Parameters for the spectrum approximation of Eq.~\eqref{eq:fit_function} for various coupling structures. 
}
	\label{fig:fit_parameters}
\end{figure*}

\begin{figure}[t]
    \includegraphics[width=0.49\textwidth]{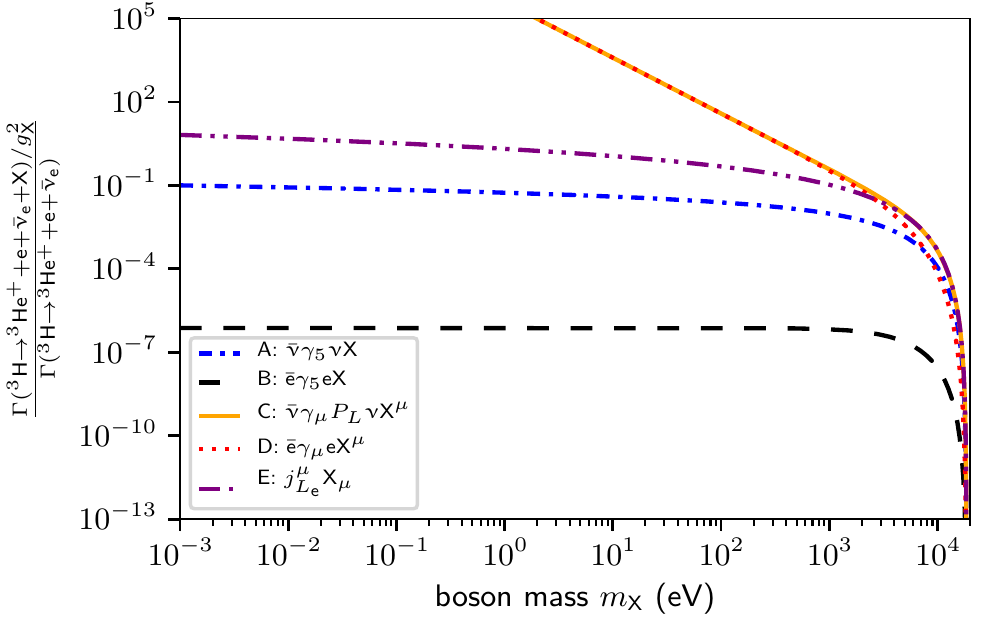}
	\caption{Ratio of the decay width 
		$\Gamma({}^3\text{H} \to {}^3\text{He}^{+} + e^-+ \bar \nu_e + X)$ (divided by the relevant coupling constant $g_X^2$) with respect to the SM width as a function of the new boson's mass $m_X$. The different curves are for different underlying coupling structures, see text for details. 
}
	\label{fig:all_branching_ratios}
\end{figure}

In the case of a pseudoscalar emitted from the outgoing neutrino we find $K\simeq 10^{-25}  g_{\nu J}^2$ and $n\simeq 2$ for $m_J$ around eV, see Fig.~\ref{fig:fit_parameters} for the full $m_J$ dependence. This implies branching ratios below $0.1 g_{\nu J}^2$ with respect to the SM decay rate, as shown in Fig.~\ref{fig:all_branching_ratios}. For larger $J$ masses the branching ratio goes down due to phase-space closure. Note that we find a small logarithmic divergence that arises for small $m_J$ due to the emission off of a massless neutrino.

\subsection{\label{sec:Pe}Pseudoscalars emitted from electrons}

As a second example, consider a pseudoscalar $J$ coupled to electrons. The relevant part of the Lagrangian takes the form 
\begin{equation} \label{eq:eeJ}
{\cal L} = i g_{eJ} \, \bar e \, \gamma_5 \, e \, J \,. 
\end{equation}
This kind of couplings typically appears in axion models or, at least at one-loop, in Majoron models~\cite{Chikashige:1980ui,Schechter:1981cv,Pilaftsis:1993af,Garcia-Cely:2017oco}. A light pseudoscalar can be also part of an extended Higgs sector. We stay agnostic about the origin of $J$ and keep its mass $m_J$ and coupling $g_{eJ}$ as free parameters to be determined experimentally.  

The resulting electron spectrum can again be extremely well described by the simple function of Eq.~\eqref{eq:approx_spectrum}. For $m_J$ below eV, the relevant parameters are $K\simeq 1.4\times 10^{-27} g_{eJ}^2$ and $n\simeq 4$. As can be seen from Fig.~\ref{fig:all_branching_ratios}, the resulting branching ratio is rather small even for $g_{eJ}\sim 1$.
Note that the amplitude is well-behaved in the limit $m_J\to 0$, unlike for the neutrino coupling. 

\subsection{\label{sec:Vnu}Vector bosons emitted from neutrinos}

An interesting possibility arises via the presence of a light neutral vector boson $Z'$. Such bosons are a common feature in many beyond-the-SM frameworks with additional gauge symmetries. Their masses are typically determined by a (combination of) gauge coupling(s) and the energy scale at which the additional gauge symmetry is spontaneously broken~\cite{Langacker:2008yv}, or in some other cases via the St{\"u}ckelberg mechanism~\cite{Ruegg:2003ps}. Either way, it is conceivable to have light $Z'$ bosons, which are subject of intensive studies, for instance if they behave like dark photons~\cite{Alexander:2016aln}. 
Recent model building perspectives for a light $Z'$ coupled to neutrinos can be found in Refs.~\cite{Farzan:2016wym,Bakhti:2018avv}. 
We will focus on an effective description of $Z'$ interactions with left-handed neutrinos as described by the Lagrangian 
\begin{equation}\label{eq:vvZ}
{\cal L} = g_{\nu L} \, \bar \nu \gamma^\mu P_L  \nu Z'_\mu \,.
\end{equation}
For $m_{Z'}$ below $\unit[100]{eV}$, the spectrum is well-described by  Eq.~\eqref{eq:approx_spectrum} with $n\simeq 4$ and $K \simeq 6.7\times 10^{-22} g_{\nu L}^2 (\unit{keV}/m_{Z'})^2$, see Fig.~\ref{fig:fit_parameters}. The decay width shows the characteristic $1/m_{Z'}^2$ behavior from the longitudinal component of the $Z'$, as expected from a coupling to a non-conserved current~\cite{Dror:2017ehi,Dror:2017nsg}. Note that this is the result of the non-renormalizability of the Lagrangian, i.e.~an effective field theory. A full UV-complete theory would eventually correct this feature, which is however a quite model-dependent effect. We assume here 
that the solution to the $1/m_{Z'}^2$ behavior does not kick in before the mass scales that \katrin~can probe. Nevertheless, the unphysical low mass behavior of this particular model (and the one in the next subsection), illustrates that a measurement at low energies gives crucial information, which is different from extrapolating limits obtained at higher energy scale. 


\subsection{\label{sec:Ve}Vector bosons emitted from electrons}
In a similar vein, we assume the following Lagrangian for the coupling of a $Z'$ with 
electrons: 
\begin{equation}\label{eq:eeZ}
{\cal L} = \bar e \gamma^\mu (g_{eL} \, P_L + g_{eR} \, P_R ) e Z'_\mu \,.
\end{equation}
In principle one can imagine different couplings for the chiralities, 
$g_{eL} \neq g_{eR}$, but we will take them to be equal for simplicity, $g_{eL}=g_{eR}\equiv g_{eV}$, corresponding to a vector-like coupling.

For small $m_{Z'}$, the spectrum is identical to that of Eq.~\eqref{eq:vvZ}, i.e.~with parameters $n\simeq 4$ and $K \simeq 6.7\times 10^{-22} g_{eV}^2 (\unit{keV}/m_{Z'})^2$, see Fig.~\ref{fig:fit_parameters}. This is not surprising upon noting that the electron-number current $j^\alpha_{L_e}=\bar \nu_e \gamma^\alpha P_L  \nu_e+ \bar e \gamma^\alpha e$ is classically conserved, so a light $Z'$ coupled to it would not lead to a $1/m_{Z'}^2$ divergence (see Sec.~\ref{sec:Le}). This merely means that the $1/m_{Z'}^2$ divergence of the $\bar e \slashed{Z'} e$ coupling is exactly canceled by the $1/m_{Z'}^2$ divergence of the $\bar \nu_e \slashed{Z'}  P_L  \nu_e$ coupling.

There are of course differences between the $Z'$ coupling to electrons and neutrinos, at least for larger $Z'$ masses where the $1/m_{Z'}^2$ behavior is less dramatic. From Fig.~\ref{fig:fit_parameters} we can see that the two couplings become distinguishable for boson masses above keV.

\subsection{\label{sec:Le}Vector bosons emitted from neutrinos and electrons}

As mentioned above, a $Z'$ that is coupled to the classically conserved electron-number current $j_{L_e}$ has a qualitatively different behavior than a $Z'$ that only couples to electrons or neutrinos. With 
\begin{equation}\label{eq:LeZ}
{\cal L} = g_{L_e} \, j^\alpha_{L_e} Z'_\alpha  = g_{L_e}\left(\bar \nu_e \gamma^\alpha P_L  \nu_e+ \bar e \gamma^\alpha e\right)Z'_\alpha 
\end{equation}
one finds that the amplitude for $Z'$ emission is approximately constant in the limit of small $m_{Z'}$, contrary to the two cases discussed above, because $j^\alpha_{L_e}$ is a classically conserved (non-anomalous) current.  Fitting to the approximate spectrum gives $n\simeq 2.2$ and $K \simeq 5\times 10^{-24} g_{L_e}^2$ for $m_{Z'}\sim \unit{eV}$, see Fig.~\ref{fig:fit_parameters}. Similar to the scalar-neutrino case there is still a logarithmic dependence on $m_{Z'}$ that is well known from Bremsstrahlung in QED. 
For larger $m_{Z'}\gtrsim \unit[10]{keV}$, the behavior becomes identical to that of a $Z'$ coupled to neutrinos.

\section{\label{sec:eV}\si{\electronvolt}-scale light bosons in the current \katrin~setup}
\noindent In order to compare the theoretical prediction for the light bosons to upcoming data taken by the \katrin~experiment in the current setting, we will apply several modifications to the analytical form given in Eq.~\eqref{eq:approx_spectrum} to account for experimental characteristics. We will refer to this as the \textit{experimental spectrum}. In the following, we will introduce the modifications to the spectrum before stating the sensitivity of \katrin~to constrain the emission of a light boson in tritium \textbeta-decay. For better readability and overview, we summarize the possible production mechanisms in Tab.~\ref{tab:ProductionMechanisms} which we will later refer to when stating the results.
\begin{table}[ht]
	\centering
    \begin{tabular}{l  c  c}
    	\toprule
		Mechanism	& $K/g_X^2$	& $n$\\
		\colrule
		A: Pseudoscalars from neutrinos	& $10^{-25}$									& 2 \\
        B: Pseudoscalars from electrons	& $10^{-27}$									& 4 \\
		C: Vector bosons from neutrinos	& $7\cdot10^{-16}(\text{eV}/m_X)^2$	& 4 \\
		D: Vector bosons from electrons	& $7\cdot10^{-16}(\text{eV}/m_X)^2$	& 4 \\
		E: Vector bosons from both		& $5\cdot10^{-24}$							& 2 \\
        \botrule
	\end{tabular}
    \caption[Different production mechanisms]{\textbf{Different production mechanisms.} This table gives an overview about the different production mechanisms for the light boson $X$ and their effect on the spectrum, caused by different forms for $K$ and $n$. The values stated are approximations for eV-scale light bosons.}
    \label{tab:ProductionMechanisms}
\end{table}

\subsection{Experimental characteristics}\label{sec:spectrum:basicform:experiment}
Since \katrin\ employs gaseous molecular tritium ($\text{T}_2$), we have to use the molecular masses to calculate the endpoint:
\begin{equation}
	E_e^{\text{max}'} = \frac{m^2_{\text{T}_2}-(m_{\text{T}\,^3\text{He}^+}+m_\upnu+m_X)^2+m_e^2}{2m_{\text{T}_2}}\,,
	\label{eq:EmaxKATRIN}
\end{equation}
which for $m_\upnu=m_X=0$ gives $E_e^{\text{max}'}=\SI{18575}{\electronvolt}+m_e$.

\begin{figure}[t]
	\centering
	\includegraphics[width=.5\textwidth]{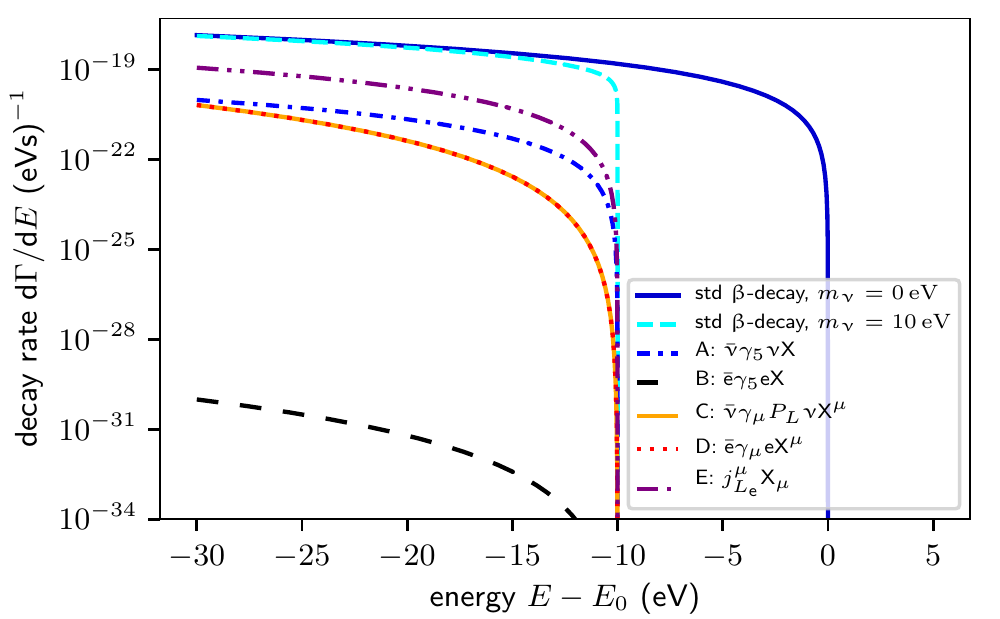}
	\caption[Differential spectrum]{\textbf{Differential spectrum} - Shown are 7 different cases: standard \textbeta-decay without (solid blue) and with \SI{10}{\electronvolt} (dashed cyan) neutrino mass, as well as the 5 spectra as expected for the additional emittance of a light boson $X$ with $g_X=1$, $m_X=\SI{10}{\electronvolt}$ with type according to Tab.~\ref{tab:ProductionMechanisms}. (Legend: dash dotted blue - A, dashed black - B, solid orange/dotted red - C/D, dash dotted dotted magenta - E.)}
	\label{fig:DiffSpec}
\end{figure}

Switching to kinetic electron energy $E=E_e-m_e$ instead of total electron energy $E_e$ results in
\begin{equation}
	\frac{\text{d}\Gamma}{\text{d}E} = \frac{K}{\hbar}\sqrt{\frac{E}{m_e}} \left(\frac{E_\text{max}-E}{E_\text{max}+m_e}\right)^n F(Z, E)\,,
	\label{eq:SpecFormKATRIN}
\end{equation}
with
\begin{align}
\begin{split}
	E_\text{max} &= E_e^{\text{max}'} - m_e - (m_\upnu + m_X) \\
   &= E_0 - (m_\upnu + m_X) \,.
\end{split}
	\label{eq:Endpoint}
\end{align}
The spectra of the 5 different cases, together with the standard \textbeta-decay spectrum for different neutrino masses, are plotted in Fig.~\ref{fig:DiffSpec}.

The detailed modeling of the decay spectrum requires several modifications to the analytical description~\cite{Kleesiek:2018mel}, leading to
\begin{align}
	\frac{\text{d}\Gamma}{\text{d}E} &= \frac{K}{\hbar}\sqrt{\frac{E}{m_e}}\, F(Z, E)\label{eq:SpecFormKATRINcorr}\\
    &\quad\times
	\sum_{fs}P_{fs}\cdot f_\text{rad}(E-E_{fs})\cdot\left(\frac{E_\text{max, fs}-E}{E_\text{max, fs}+m_e}\right)^n\nonumber,
\end{align}
whose factors will be discussed briefly in this section.

\paragraph{Radiative corrections:}\label{sec:spectrum:radiative}Due to interaction with virtual and soft real photons in the Coulomb field of the nucleus, the emitted electrons lose energy. This loss is accounted for by the correction factor $f_\text{rad}(E-E_{fs})$ in Eq.~\eqref{eq:SpecFormKATRINcorr}, as recommended in Ref.~\cite{Repko:1984cs}.

\paragraph{Molecular recoil:}\label{sec:spectrum:recoil}
As discussed in the \katrin~design report~\cite{Angrik:2005ep}, in the region around the endpoint the electron energy dominates over the neutrino energy. Therefore, the recoil energy of the molecule balances the momentum of the electron:
\begin{equation}
	E_\text{rec} \approx E\cdot\frac{m_e}{m_{^3\text{HeT}^+}}\,.
	\label{eq:recoil}
\end{equation}
We include this \SI{1.7}{\electronvolt} shift into the final states distribution~\cite{Kleesiek:2018mel}.

\paragraph{Final states:}\label{sec:spectrum:corrections:finalstates}
\katrin~is using a molecular tritium source, which contains several tritiated hydrogen isotopologues. Those are T$_2$, DT and HT which decay into ($^3$HeT)$^+$, ($^3$HeD)$^+$ and ($^3$HeH)$^+$. The dominant isotopologue 
will be T$_2$, due to the tritium purity $\epsilon_\text{T}>0.95$. The decay may leave the daughter molecule in a rovibronic (rotational and vibrational) or electronic excited final state. Distributions of these excited states were calculated by Saenz and others~\cite{Saenz:2000dul,Doss:2006zv,Doss2008} and are quantified in terms of excitation energy $E_{fs}$ and the corresponding probability $P_{fs}$. 
The final states energy $E_{fs}$ reduces the maximum energy of the electron: Eq.~\eqref{eq:Endpoint} thus becomes
\begin{equation}
	E_{{\rm max}, fs} = E_0 - E_{fs} - (m_\upnu + m_X) \,,
	\label{eq:FinalStates}
\end{equation}
requiring the summation of the decay rate over all possible final states $fs$ in Eq.~\eqref{eq:SpecFormKATRINcorr}.


\paragraph{Doppler effect:}\label{sec:spectrum:corrections:doppler}
The tritium molecules in the source are at a non-zero temperature of \SI{30}{\kelvin}, which causes thermal motion. This thermal motion together with the bulk velocity of the gas flow is called Doppler effect and causes a Gaussian broadening of the electron energy spectrum of about \SI{100}{\milli\electronvolt}~\cite{Kleesiek:2018mel}.


\subsection{Combination with standard \textbeta-decay spectrum}\label{sec:spectrum:combination}
\begin{figure}[t]
	\centering
	\includegraphics[width=.5\textwidth]{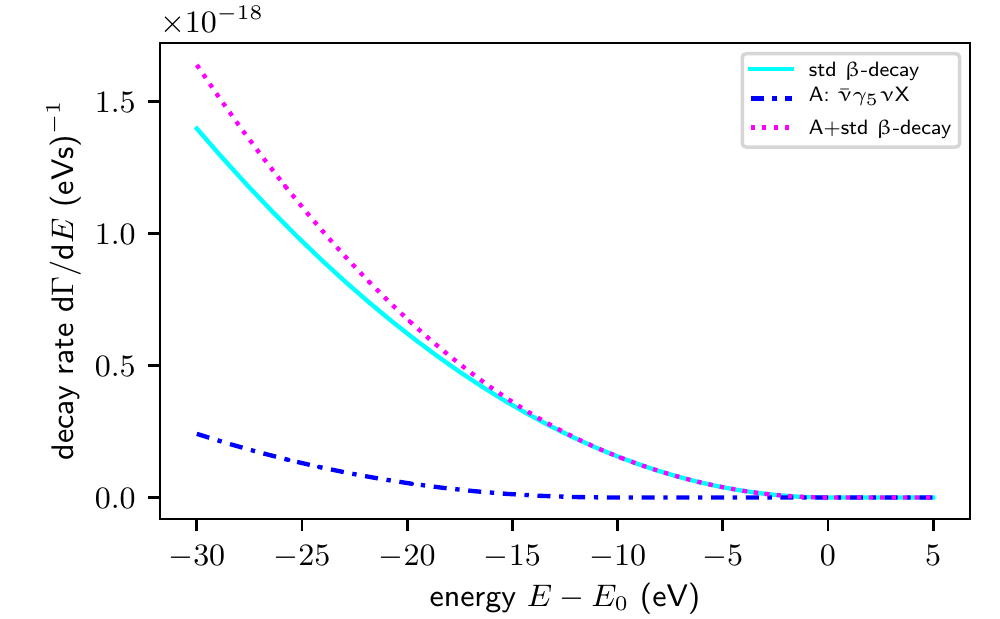}
	\caption[Combination with standard \textbeta-decay]{\textbf{Combination with standard \textbeta-decay} - Shown is the signature of a light boson of type A with $g_{\nu J}=5$, $m_J=\SI{10}{\electronvolt}$ together with standard \textbeta-decay spectrum with vanishing neutrino mass.}
	\label{fig:CompSpectra}
\end{figure}

We define the form of the overall spectrum as
\begin{equation}
\frac{\text{d}\Gamma}{\text{d}E} = \left.\frac{\text{d}\Gamma}{\text{d}E}\right|_\upbeta + \left.\frac{\text{d}\Gamma}{\text{d}E}\right|_X.
\label{eq:SpecCombNoNormalisation}
\end{equation}
Fig.~\ref{fig:CompSpectra} shows the superposition of a light boson of type~A (large coupling $g_{\nu J}=5$ assumed for visualization) for $m_J=\SI{10}{\electronvolt}$ and of the standard \textbeta-spectrum with vanishing neutrino mass. Note that the common normalization parameter ensures that the decay activity stays constant.

\katrin~measures an integrated spectrum with the high voltage $U$ at the main spectrometer acting as a high-pass filter~\cite{Angrik:2005ep}. Using the concept of a response function $R(E, qU)$~\cite{Angrik:2005ep,Kleesiek:2018mel}, the measured spectrum can be written as
\begin{equation}
	\dot{N} \propto \int_{qU}^{\infty} R(E, qU) \left(\left.\frac{\dd\Gamma}{\dd E}\right|_\upbeta + \left.\frac{\dd\Gamma}{\dd E}\right|_X\right) \dd E\,.
	\label{eq:IntSpec}
\end{equation} 
\subsection{Settings}\label{sec:settings}

\begin{figure}[ht]
	\centering
	\includegraphics[width=.5\textwidth]{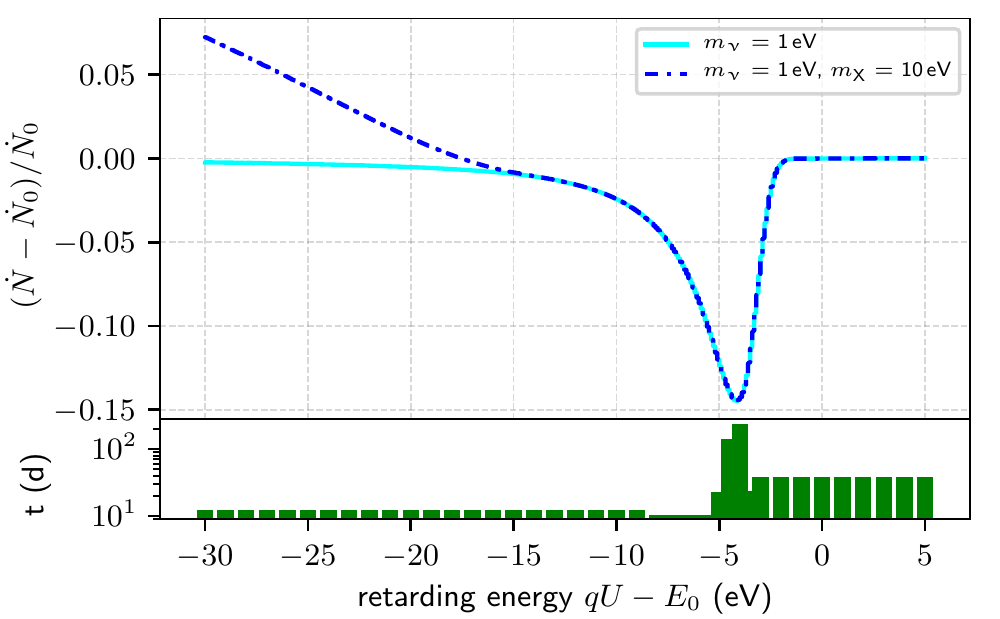}
	\caption[Relative change on integrated spectrum with MTD]{\textbf{Relative change of the integrated spectrum} - Shown is the neutrino mass sensitive region by comparing a spectrum with neutrino mass $m_\upnu=\SI{1}{\electronvolt}$ to a spectrum without neutrino mass (solid cyan). The peak of the measuring time distribution  sits right in the region where the neutrino mass causes the largest distortion to the spectrum. Furthermore the signal of a light boson type A with $g_{\nu J}=5$, $m_X=\SI{10}{\electronvolt}$ additional to the massive neutrino is compared to the null-hypothesis (dash dotted blue).}
	\label{fig:LikelihoodSpectra}
\end{figure}


The light boson spectrum superimposed to the standard \textbeta-decay spectrum has 6 fit parameters:
\begin{enumerate}
	\item neutrino mass squared $m_\upnu^2$
	\item endpoint $E_0$
    \item amplitude $Amp$ which is a factor ensuring correct normalization of the superposition of light boson and standard \textbeta-decay spectrum
	\item background rate $Bg$
	\item light boson mass $m_X$
	\item light boson coupling $g_X$
\end{enumerate}
In the following we use the standard settings defined in the design report~\cite{Angrik:2005ep}. This includes a measuring window of high voltage values ranging in the interval $[E_0-\SI{30}{\electronvolt}, E_0+\SI{5}{\electronvolt}]$ (see Fig.~\ref{fig:LikelihoodSpectra}), a true neutrino mass of zero and a measuring time of three years.

\subsection{Statistical sensitivity}\label{sec:sensitivity}
Here we state statistical sensitivity estimates for the potential of \katrin~to constrain the emittance of a light boson additional to the standard \textbeta-decay. Estimation of confidence intervals in the presence of nuisance parameters $\pi$ can easily lead to errors, for instance if one neglects correlations between the nuisance parameters and the parameters of interest $\Theta$. Thereby, errors on parameters of interest may be underestimated.

To minimize this risk, we will make use of the so-called profile likelihood method~\cite{Rolke:2004mj}. Using a likelihood ratio test statistic which converges to a $\chi^2$ random variable, we can extract confidence limits from the likelihood function similar to the $\chi^2$ method.
We define the profile likelihood
\begin{equation}
	L_p(\Theta)=L(\Theta,\hat{\pi}(\Theta))
	\label{eq:ProfLL}
\end{equation}
with $\hat{\pi}(\Theta)$ being the function that maximizes the likelihood with respect to its nuisance parameters $\pi$. The profile likelihood therefore only depends on the parameters of interest $\Theta$. Using the best-fit estimate $\hat{\Theta}$, we define the likelihood ratio test statistic
\begin{equation}
	\lambda(\Theta) = \frac{L_p(\Theta)}{L_p(\hat{\Theta})}\,.
	\label{eq:LikelihoodRatio}
\end{equation}
Now we can scan the profile likelihood to find the values of $\Theta$ where Eq.~\eqref{eq:ProfLL} increases by a specific factor. For example to find the $1\sigma$ intervals of a single parameter $\Theta$ one would search for $\Theta$ where $-2\Delta\log L_p(\Theta) = -2\log\lambda=1$.

\begin{figure}[t]
	\centering
    \includegraphics[width=.5\textwidth]{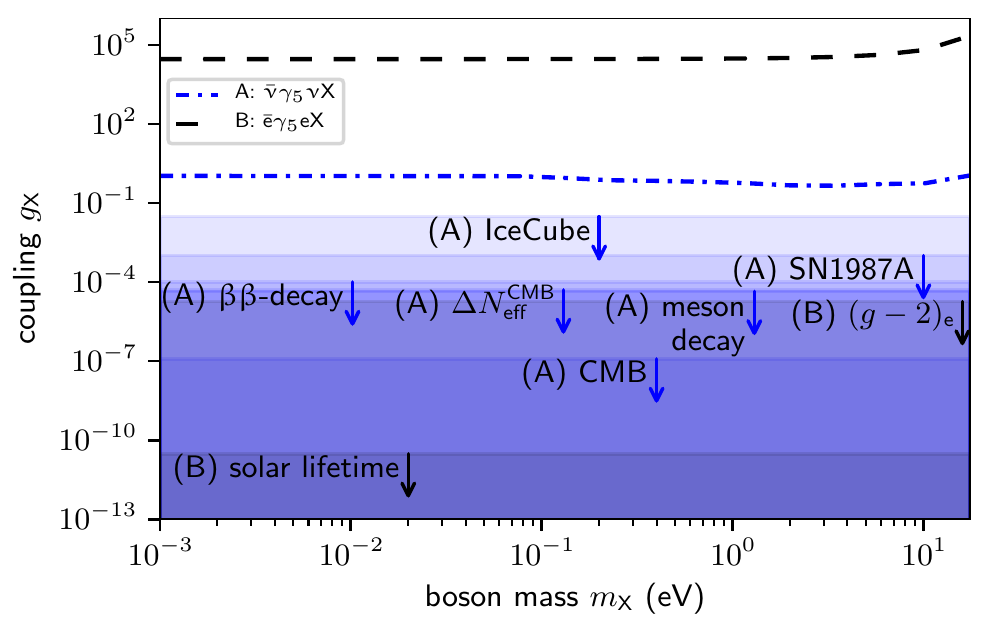}
	\caption[Statistical sensitivity contours for eV-scale light pseudoscalar bosons]{\textbf{Statistical sensitivity contours for eV-scale light pseudoscalar bosons} - Shown is the 90\,\%\,C.L.\ statistical sensitivity of \katrin~for the detection of eV-scale light pseudoscalar bosons. The types A, B are defined according to Tab.~\ref{tab:ProductionMechanisms}. Furthermore, the shaded areas mark the parameter regions allowed from constraints discussed in Sec.~\ref{sec:other}.}
	\label{fig:SensitivityPseudoscalar-eV}
\end{figure}

\begin{figure}[t]
	\centering
    \includegraphics[width=.5\textwidth]{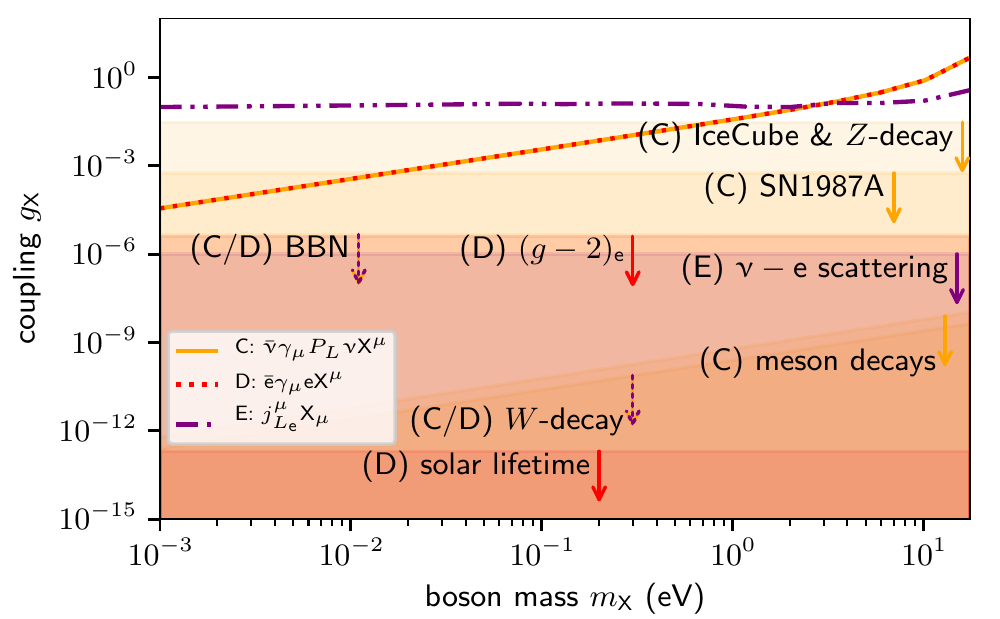}
	\caption[Statistical sensitivity contours for eV-scale light vector bosons]{\textbf{Statistical sensitivity contours for eV-scale light vector bosons} - Shown is the 90\,\%\,C.L.\ statistical sensitivity of \katrin~for the detection of eV-scale light vector bosons. The types C, D, E are defined according to Tab.~\ref{tab:ProductionMechanisms} (C and D are indistinguishable for eV-scale vector bosons). Furthermore, the shaded areas mark the parameter regions allowed from constraints discussed in Sec.~\ref{sec:other}.}
	\label{fig:SensitivityVectorboson-eV}
\end{figure}

In our case, we have two parameters of interest, namely the coupling $g_X$ and the mass of the light boson $m_X$. To determine the sensitivity, we compare our likelihood for  non-vanishing $g_X$ and $m_X$ against the null-hypothesis of no light boson: for every point in the 2-dim grid of ($g_X$, $m_X$) we minimize the likelihood with respect to the nuisance parameters $m_\upnu^2$, $E_0$, $Amp$ and $Bg$. We then can find the likelihood ratios corresponding to 90\,\%\,C.L. The resulting sensitivity curve for pseudoscalar bosons is shown in Fig.~\ref{fig:SensitivityPseudoscalar-eV} and for vector bosons in Fig.~\ref{fig:SensitivityVectorboson-eV}.
%


\paragraph*{(A) Pseudoscalars emitted from neutrinos:}\label{sec:sensitivity:PseudoscalarsNeutrinos}
Here we expect best sensitivity with increasing coupling $g_{\nu J}$ due to the form of the spectrum (compare Tab.~\ref{tab:ProductionMechanisms} and Eq.~\eqref{eq:SpecFormKATRIN}): the factor $K$ is proportional to $g_{\nu J}^2$. As can be seen from Fig.~\ref{fig:SensitivityPseudoscalar-eV}, this expectation is confirmed. Also we can see that the sensitivity decreases again for masses larger than \SI{10}{\electronvolt}: the mass limit is constrained by the extent of the measuring time \katrin~is using to scan the spectrum. Masses larger than \SI{30}{\electronvolt} are not accessible in this study due to the used measuring time distribution (compare Fig.~\ref{fig:LikelihoodSpectra}), however, there is a small increase in sensitivity for boson masses around \SI{1}{\electronvolt}.

\paragraph*{(B) Pseudoscalars emitted from electrons:}\label{sec:sensitivity:PseudoscalarsElectrons}
As mentioned in the beginning, this production mechanism for the light bosons is expected to be suppressed compared to the others due to the small branching ratio. Nevertheless Fig.~\ref{fig:SensitivityPseudoscalar-eV} shows the expected sensitivity of \katrin~towards this light boson type B. As expected, large couplings $g_{eJ}$ are required in order for \katrin~to be sensitive towards this kind of boson.


\paragraph*{(C, D) Vector bosons emitted from neutrinos or electrons:}\label{sec:sensitivity:VectorbosonsNeutrinosOrElectrons}
For the production mechanisms C and D (which electron spectra look exactly the same for eV-scale bosons), we expect an inverted behavior for low couplings compared to case A: electron spectra coupled to bosons of type C and D have a $1/m_{Z^{'}}^2$ divergence in their $K$. Therefore, small boson masses are strongly favored in this case. This should lead to higher sensitivity of \katrin~for smaller boson masses. Exactly this behavior can be seen from Fig.~\ref{fig:SensitivityVectorboson-eV}: C and D in contrast to A and B have good sensitivity towards lower boson masses.

\paragraph*{(E) Vector bosons emitted from neutrinos and electrons:}\label{sec:sensitivity:VectorbosonsNeutrinosAndElectrons}
For production mechanism E, we have spectrum parameters similar to A (compare Tab.~\ref{tab:ProductionMechanisms}) and therefore expect a similar sensitivity curve. Fig.~\ref{fig:SensitivityVectorboson-eV} confirms this expectation: the best sensitivity is expected for a light boson mass around \SI{1}{\electronvolt}.

It has to be noted that we conducted this study for each type of light boson separately. In the final analysis, only the most physically relevant case might be considered. Furthermore, we want to stress that this study might additionally be evaluated with possible eV-sterile neutrino mass and mixing angle as additional nuisance parameters. It was also checked that moderate constraints on the coupling of the light bosons are enough to not impact the \katrin~neutrino mass sensitivity in the light boson scenario at hand. For example, a conservative constraint on the coupling of boson type~E to values below $1$ preserves the \katrin~neutrino mass sensitivity of \SI{200}{\milli\electronvolt}.
In order to derive the final experimental sensitivity for the different cases, systematic effects as described in~\cite{Angrik:2005ep} need to be evaluated with respect to each specific spectrum.

\section{\si{\kilo\electronvolt}-scale light bosons: statistical sensitivity}
\label{sec:keV}
\noindent Following the study in Ref.~\cite{Mertens:2014nha} we consider a measurement of the complete differential \textbeta{}-decay spectrum at the \katrin~experiment with a new detector and readout system.
In this case it is the detector itself which determines the electron energy.
The main spectrometer is kept at a small retarding potential to allow electrons from most of the spectrum to reach the detector.
A new detector system is needed to handle much higher count rates in the whole spectrum as compared to the endpoint region, provide better energy resolution, and limit several systematic effects that arise when measuring the full \textbeta{}-spectrum.

As a case study of expected statistical sensitivity to the light boson coupling constant $g_X$ and mass $m_X$ we consider the differential measurement with the design \katrin~setup but modified detector system for a duration of three years.
The corresponding total statistics amounts to about $10^{18}$ electrons.
Furthermore, we assume for the energy resolution a conservative full width at half maximum of \SI{300}{\electronvolt}, based on recent evaluation of the detector prototype \cite{Mertens:2018}, and constant background rate of \SI{2}{\milli\cps\per\kilo\electronvolt}, based on measurements with the existing detector at the \katrin~setup \cite{Harms:2015}.
Evidently, for \si{\kilo\electronvolt}-scale boson the normalization prefactor $K$ and spectral index $n$ are no longer constants and must be considered as functions of the boson mass, see Fig.~\ref{fig:fit_parameters}.

\begin{figure}[t]
  \centering
  \includegraphics[width=1.0\columnwidth]{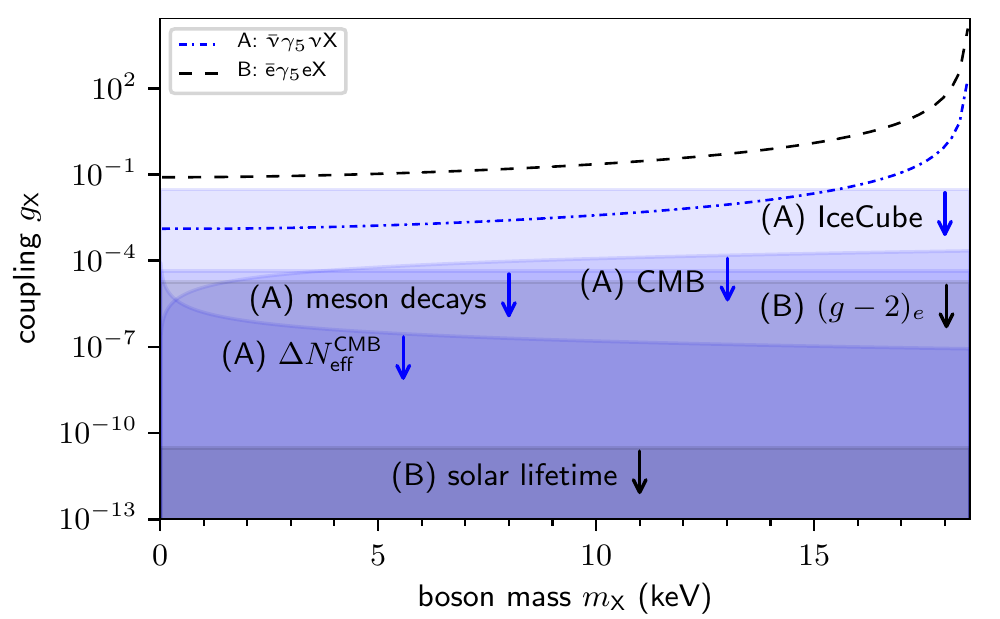}
  \caption{\textbf{Purely statistical sensitivity contours for \si{\kilo\electronvolt}-scale light pseudoscalar bosons} - Shown is the  \SI{90}{\percent}~C.L.\  statistical sensitivity of a \katrin{} differential spectrum measurement with a new detector system. Furthermore, the shaded areas mark the parameter regions allowed from constraints discussed in Sec.~\ref{sec:other}.}
  \label{fig:sensitivity_keV_pseudoscalar}
\end{figure}

\begin{figure}[t]
  \centering
  \includegraphics[width=1.0\columnwidth]{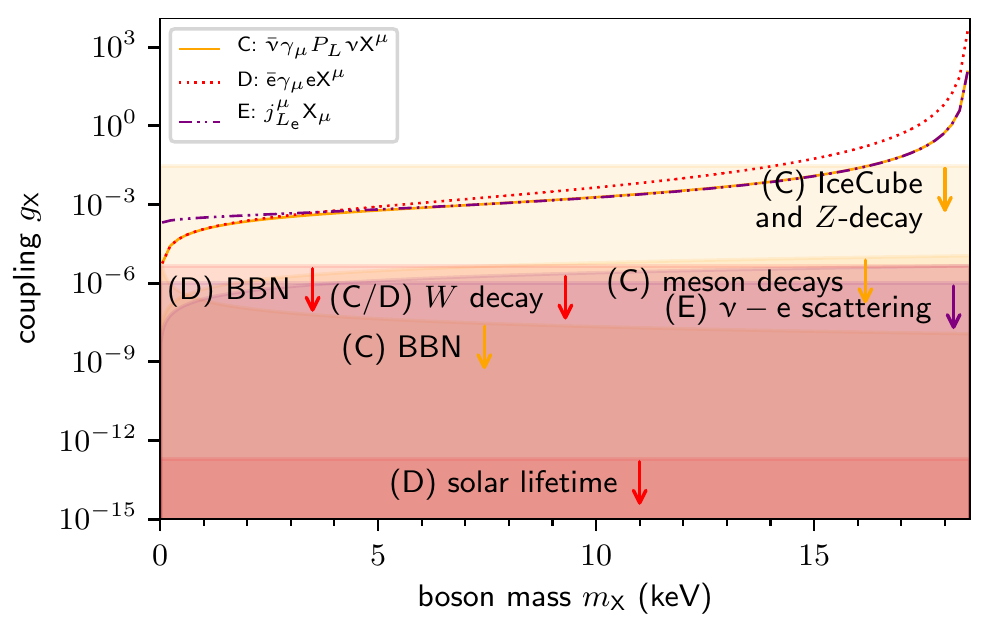}
  \caption{\textbf{Purely statistical sensitivity contours for \si{\kilo\electronvolt}-scale light vector bosons} - Shown is the  \SI{90}{\percent}~C.L.\  statistical sensitivity of a \katrin{} differential spectrum measurement with a new detector system. Furthermore, the shaded areas mark the parameter regions allowed from constraints discussed in Sec.~\ref{sec:other}.}
  \label{fig:sensitivity_keV_vector}
\end{figure}

Using the profile likelihood method similarly as in Sec.~\ref{sec:eV}  we obtain the \SI{90}{\percent}~C.L.\ sensitivity curves in Figs.~\ref{fig:sensitivity_keV_pseudoscalar} and~\ref{fig:sensitivity_keV_vector}.
In the study we have assumed a Gaussian pull term on the neutrino mass with a width of \SI{2}{\electronvolt}.
The sensitivity drops significantly for larger boson masses as expected due to the decreasing decay width of the light boson relative to the SM width.
As in the \si{\electronvolt}-scale case the branching ratio is generally small for the pseudoscalar emitted from electrons (type B), leading to relatively worse sensitivity with respect to the other production mechanisms.
For vector bosons emitted either from neutrinos (C) or electrons (D) the sensitivity increases significantly for low masses due to the $1/m_{Z^{'}}^2$ behavior of the decay width.
Furthermore, the similar experimental sensitivity reflects the similarity of the spectrum of vector boson emitted from electrons (D) to that from neutrinos (C) for small masses as well as the spectrum of vector boson emitted from both electrons and neutrinos (E) to that from neutrinos (C) for larger masses.

The huge statistics available from tritium \textbeta{}-decay in a \katrin-like experiment allows probing the \si{\kilo\electronvolt}-scale boson coupling constant to as low as $g_X^2 \sim 10^{-8}$ for some of the models.
Nevertheless, as already recognized for the \si{\kilo\electronvolt}-scale sterile neutrino in Ref.~\cite{Mertens:2018} the final experimental sensitivity will be limited by unavoidable systematic effects connected to observing the entire \textbeta{}-spectrum.
Detailed studies are thus required in order to assess the final experimental sensitivity.
Besides, the presented statistical sensitivity is valid when the experimental search is done for a given most physically relevant type of particle.

\section{\label{sec:other} Additional bounds on light states}
\noindent
New light states are actively searched for at laboratory scales also through different processes with respect to the one considered in this work. Moreover, their interactions sensibly affect cosmological and astrophysical processes. This leads to potentially very strong bounds on the strength of their interaction affecting the expected sensitivity region for the \katrin~experiment. Below we briefly illustrate and discuss the most relevant bounds for the scenarios under consideration.\\

Concerning possible complementary or competitive laboratory constraints, we have first of all to consider the emission of light bosons in the decays of the $Z$, $W$, and of light charged mesons $P$, hence leading to three-body decay processes like $Z \rightarrow \nu \nu X$, $W \rightarrow e \nu X$ and $P \rightarrow e \nu X$. Three-body decay rates of the $Z$ and $W$ bosons are strongly constrained by the very precise measurements of their total decay widths, while the decay rate of the mesons are probed by dedicated searches. In the case of models C and E we consider the bound reported in~\cite{Laha:2013xua} on the $Z \rightarrow \nu \nu X$ process, which can be approximately\footnote{The rate $\Gamma(Z \rightarrow \nu \nu Z^{'})$ depends only logarithmically on the mass of the light vector.} expressed as $g_{\nu L}, g_{L_e} \lesssim 3 \times 10^{-2}$.
Concerning $W$ decay, the cases of models $C/D$ and $E$ are very different. For the former, similarly to what occurs in the model presented in~\cite{Laha:2013xua}, the $Z^{'}$ is coupled to an anomalous current of SM leptons, leading to a $1/m_{Z'}^2$ enhancement of the three body decay rate of the $W$, determining the very strong bound:
\begin{equation}
g_{eV},g_{\nu L} \lesssim 2.5 \times 10^{-7} \left(\frac{m_{Z'}}{\unit[1]{keV}}\right) .
\end{equation}
Note that we need to extrapolate this bound to low masses in order to apply it to our scenario. 
Model E, on the contrary, does not feature this anomalous coupling, hence the decay rate of $W \rightarrow e \nu Z^{'}$ depends only logarithmically on $m_{Z'}$ and gives a weaker limit as the one from $Z \rightarrow \nu \nu Z^{'}$.
Bounds from light meson decays applicable to model C have been considered in~\cite{Bakhti:2017jhm}. For our study we will adopt the bound from the process\footnote{We have conservatively adopted the bound from $\pi \rightarrow e \nu Z^{'}$ rather than the slightly stronger one from $K \rightarrow e \nu \nu \nu$ since the former is independent of the lifetime of the $Z^{'}$.} $\pi \rightarrow e \nu Z^{'}$:
\begin{equation}
\label{eq:meson_bound}
g_{\nu L} \lesssim 6 \times 10^{-7} \left(\frac{m_{Z'}}{\unit[1]{keV}}\right) .
\end{equation}
Slightly stronger bounds would be obtained from violation of lepton universality~\cite{Bakhti:2017jhm,Bakhti:2018avv}. These would be, however, dependent on eventual couplings of the light vector with second generation leptons. 
According to an analogous reasoning as for the case of $W$ decays, Eq.~(\ref{eq:meson_bound}) cannot be applied to model E, since the $1/m_{Z'}^2$ enhancement of the decay rate would not be present. 

Limits applicable to model A from decays of light mesons have been provided in~\cite{Pasquini:2015fjv} (see also~\cite{Lessa:2007up,Albert:2014fya}). For our analysis we will adopt the constraint $g_{\nu J} < 4.4 \times 10^{-5}$, independent on the mass of the pseudoscalar $J$ for the whole range of masses considered in this work.


Models B, D and E are also constrained by the determination of the anomalous magnetic moment of the electron~\cite{Tanabashi:2018oca,Parker191}. The corresponding upper bounds read~\cite{Lindner:2016bgg,Liu:2018xkx}:
\begin{align}
\begin{split}
& g_{eJ} \lesssim 1.8 \times 10^{-5} \,,\\
& g_{eV} \lesssim 4.0 \times 10^{-6} \,
\end{split}
\end{align}
for the cases of a pseudoscalar (model B) and vector (models D and E) boson, respectively. As evident, the limit is stronger in the case of vector bosons. This is due to the fact that the corresponding contribution to $(g-2)_e$ adds to the SM one, in tension with the slight experimental evidence towards a negative deviation of the anomalous magnetic moment of the electron with respect to the SM expectation. 

In scenario E a strong bound $g_{eV}\lesssim 10^{-6}$~\cite{Laha:2013xua,Lindner:2018kjo} arises from electron-neutrino scattering. 

We finally remark that, in the case of a lepton number violating coupling,  constraints from neutrinoless double beta decay, determined in Majoron models, can be applied to model A. These can be expressed as an upper bound on the coupling $g_{\nu J}$ of the order $10^{-4}$ to $10^{-5}$~\cite{Blum:2018ljv,Brune:2018sab}. \\

Together with these laboratory limits, cosmological and astrophysical constraints should be considered as well.

Concerning astrophysical constraints, very strong ones apply to models B and D. Indeed, a boson $X$ coupled to an electron pair would be responsible for energy loss in the interior of stars as a consequence of Bremsstrahlung ($e+Z e \rightarrow Z e +e +X$) and Compton ($\gamma +e \rightarrow e+ X$) processes. The corresponding cross sections have been determined for the models~B and D, for example, in~\cite{Gondolo:2008dd} (this bound has been also revised in~\cite{Redondo:2008aa,An:2013yfc} for dark photon models,  potentially applicable to model D). The requirement of not altering the properties of the Sun translates into the strong constraint $g_{eJ} \lesssim 3 \times 10^{-11}$ for the case of a boson with pseudoscalar coupling to electrons (model B) and an even stronger upper bound $g_{eV} \lesssim 2\times 10^{-13}$ in the case of a boson with vectorial coupling (model D). Notice that this bound implicitly assumes that the new state is capable of escaping the Sun. It is conceivable that, for strong enough coupling, it gets instead trapped within the Sun and reprocessed back into SM states without causing energy loss. An assessment of this effect would, however, require a dedicated study which is beyond the scope of this paper.

Bounds on the interactions of a new light state with neutrinos come from the observed flux of neutrinos from galactic and extragalactic sources. In the presence of a light mediator, high-energy or ultra-high energy neutrinos would feature an enhanced scattering rate on the Cosmic Neutrino Background (C$\nu$B) implying reduced fluxes, with respect to SM expectations, at Earth. Along this way of reasoning the pioneering work~\cite{Kolb:1987qy} provided limits on the interactions of the neutrinos with a light boson based on the observation of neutrinos from the supernova SN1987A. This reference considered the cases of a very light spin-1 mediator (relevant for eV states in model C), a massive spin-1 mediator (keV states in model C) and a Majoron (relevant for eV states in model A) obtaining, respectively, the limits $g_{\nu L} \lesssim 5.6 \times 10^{-4}$, $g_{\nu L}/m_{Z'} < 12/\unit{MeV}$ and $g_{\nu J} \lesssim 10^{-3}$. According to a similar logic, Refs.~\cite{Ng:2014pca,Ioka:2014kca} considered the case of the flux of extragalactic neutrinos measured by IceCube. In particular Ref.~\cite{Ioka:2014kca} provided the limit $g_{X} \lesssim 0.03$ which can be applied to both models A and C for all the ranges of masses considered in our study. 

New light states coupled to neutrinos can also be efficiently produced in core-collapse supernovae environments and affect their evolution. Since the observed flux of neutrinos from SN1987A was compatible with standard predictions, it is possible to obtain constraints on the coupling of the new states with neutrinos. Studies along this line have been performed mostly in the context of Majoron models, see e.g.~\cite{Gelmini:1982rr,Choi:1987sd,Berezhiani:1989za,Choi:1989hi,Chang:1993yp,Kachelriess:2000qc,Tomas:2001dh,Lindner:2001th,Hannestad:2002ff,Farzan:2002wx,Fogli:2004gy,Das:2011yh,Brune:2018sab}; extensions to more generic scalar and pseudoscalar states have been considered in~\cite{Heurtier:2016otg,Farzan:2018gtr}. As a consequence of this we will consider supernova bounds in the context of model A.\footnote{An analogous reasoning could be applied also to the case of a mediator coupled to electrons. Studies along these lines have been conducted e.g.~in~\cite{Chang:2016ntp,Chang:2018rso} providing limits in the context of dark photon models.}

The environment of supernova cores is affected in different ways by the presence of  light BSM states interacting with neutrinos. First of all, the production process $\nu \nu \rightarrow J$ leads to energy depletion, hence reducing the neutrino flux, and to deleptonization (i.e.~reduction of the electron lepton number inside the supernova core), eventually preventing the explosion of the supernova. These two effects should be considered, however, only if the interactions of the state $J$ are feeble enough that it can escape the supernova core. On the contrary, for $g_{\nu J} \gtrsim 10^{-5}$, the scattering processes $\nu +J \rightarrow \nu+J$ are efficient enough to trap the $J$ particles inside the supernova core, so that they eventually decay back into neutrinos  such that no energy loss or deleptonization occurs. As evident from Figs.~\ref{fig:SensitivityPseudoscalar-eV} and~\ref{fig:sensitivity_keV_pseudoscalar} the region of interest of \katrin~lies in the trapping regime for the state $J$ and hence the constraints just mentioned do not apply in our scenario. \\

Moving finally to cosmological constraints, light states with sizable interactions with neutrinos can contribute to the effective number of the neutrino species, $N_{\rm eff}$, which is constrained both by Big Bang Nucleosynthesis (BBN) and the Cosmic Microwave Background (CMB). BBN bounds are relevant only in the case of a vector boson; a (pseudo)scalar state, even if fully in thermal equilibrium at the time of BBN, would contribute to the effective number of neutrinos by an amount of at most $\Delta N_{\rm eff} \approx 0.57$, well compatible with the bound $\Delta N_{\rm eff} <1$~\cite{Mangano:2011ar}. On the contrary a vector boson, because of the larger number of degrees of freedom, can contribute up to $\Delta N_{\rm eff}=1.71$. A simple estimate of the BBN bound, for model C, can be obtained by requiring that the relevant interaction rate of the new state with the neutrinos is below the Hubble expansion rate at the typical temperature of BBN, $T \sim \unit[1]{MeV}$. For a keV mass state the most relevant interaction is the inverse decay $\nu \nu \rightarrow Z'$. In such a case the BBN constraint is translated into an upper bound on the coupling of the form~\cite{Ahlgren:2013wba,Huang:2017egl}
\begin{equation}
g_{\nu L} \lesssim 2.2 \times 10^{-7} \left(\frac{\unit[1]{keV}}{m_{Z'}}\right) .
\end{equation} 
In the case of an eV state one should instead consider the neutrino annihilation processes $\nu \nu \rightarrow Z' Z'$ which yields the bound~\cite{Huang:2017egl}
\begin{equation}
\label{eq:BBNlimit_lowmass}
g_{\nu L} \lesssim 4.6 \times 10^{-6}\,.
\end{equation}
While these estimates are already a very good approximation, we have adopted the more refined limits determined in~\cite{Huang:2017egl} for our analysis. 
A similar way of reasoning could also be  applied for the limits on measurements of $N_{\rm eff}$ at the CMB time. The stronger constraint $\Delta N_{\rm eff} < 0.3$ at $95\%$\, C.L.~\cite{Aghanim:2018eyx} allows to probe also model C. As discussed for example in~\cite{Boehm:2012gr}, a quantitative assessment on the contribution to $\Delta N_{\rm eff}$ depends on the details of the decoupling of the light state from neutrinos and is beyond the purpose of this work. In Figs.~\ref{fig:SensitivityPseudoscalar-eV} and~\ref{fig:sensitivity_keV_pseudoscalar} we report the contour corresponding to the case of decoupling at temperatures of the order of MeV, corresponding, as already pointed out, to $\Delta N_{\rm eff}^{\rm CMB}\simeq 0.57$, which is in $2 \sigma$ tension with the experimental limit. 

Analogous bounds to the ones just discussed can be applied to models B and D. In such a case the eventual equilibration of the new boson $X$ with the thermal SM bath is mostly determined by the rates of $XX \leftrightarrow e^+ e^-$ processes. The condition for the equilibration of this rate is described in a good approximation by Eq.~(\ref{eq:BBNlimit_lowmass}). We notice, anyway, that due to the kinematical suppression of the $XX \rightarrow e^+ e^-$ rate from the mass of the electron, the $X$ states would decouple at latest at temperatures of the order of the mass of the electron, $\unit[0.5]{MeV}$.

Turning to the CMB, new light states would enhance neutrino self-interactions, implying an alteration of their free-streaming length which would result in an enhancement of the CMB temperature power spectrum at multipoles  $l \gtrsim 200$~\cite{Hannestad:2004qu,Bell:2005dr,Cyr-Racine:2013jua}. This effect is customarily analyzed in two limiting regimes. The first one holds if the mass of the mediator is significantly above the typical energy of neutrinos at CMB time (which is around eV). In such a case the only relevant processes are neutrino self-scattering processes, $\nu \nu \rightarrow \nu \nu$, which can be described through an effective four-fermion interaction with coupling $G_X=g_X^2/m_X^2$. Constraints on these effective couplings have been determined in~\cite{Cyr-Racine:2013jua,Archidiacono:2013dua} and more recently revised in~\cite{Lancaster:2017ksf,Oldengott:2017fhy}. As can be easily argued, the case of a keV state falls in this regime. We will apply the following bound, for both models A and C~\cite{Archidiacono:2013dua}:  
\begin{equation}
G_X \leq 2.5 \times 10^7 \, G_F \,,
\end{equation}
($G_F$ being the Fermi constant), which can be re-expressed as:
\begin{align}
& g_X \lesssim 1.2 \times 10^{-2} \left(\frac{m_X}{\unit[1]{MeV}}\right).
\end{align}
The second regime holds for a mediator which can be regarded as massless with respect to the energy of neutrinos at CMB. In such a case a larger variety of processes, including also neutrino annihilations into mediator pairs, should be considered. A study along these lines has been presented e.g.~in Ref.~\cite{Archidiacono:2013dua} (see also~\cite{Forastieri:2015paa}) in the case of a light pseudoscalar (model A) and the very strong bound $g_{\nu J} \lesssim 1.2 \times 10^{-7}$ has been obtained.

From the discussion above it is evident that the case of an eV state is troublesome since it does not fit any of these two regimes, since its mass of the same order as the energy of neutrinos at CMB. To our best knowledge, no limit is available for an ${\cal O}(\unit{eV})$ mass boson. \\



\begin{table*}
\renewcommand{\baselinestretch}{1.4}\normalsize 
\begin{tabular}{c|c}
\hline\hline
{\bf Model A: $i g_{\nu J}\, \bar \nu \, \gamma_5 \, \nu \, J$} & \\
\hline
Double $\beta$ decay (only LNV) & $g_{\nu J}\lesssim 10^{-(4 \div 5)}$~\cite{Blum:2018ljv,Brune:2018sab} \\
\hline
Meson decays & $g_{\nu J} \lesssim 4.4 \times 10^{-5}$~\cite{Pasquini:2015fjv}\\
\hline
CMB & $g_{\nu J} \lesssim 1.2 \times 10^{-2} \left(\frac{m_J}{\unit[1]{MeV}}\right)\,\,\left(m_J \gg \unit[1]{eV}\right)$\\
 & $g_{\nu J} \lesssim 1.2 \times 10^{-7}\,\,\left(m_J \ll \unit[1]{eV}\right)$~\cite{Archidiacono:2013dua}\\
\hline 
Supernova 1987A & $g_{\nu J} \lesssim 10^{-3}\,\,\left(m_J \leq {\cal O}(\unit[1]{eV})\right)$~\cite{Kolb:1987qy}\\
\hline
IceCube & $g_{\nu J} \lesssim 0.03$~\cite{Ioka:2014kca} \\
\hline
$\Delta N_{\rm eff}^{\rm CMB}$ & $g_{\nu J} \lesssim 1.6 \times 10^{-6} \left(\frac{\unit[1]{keV}}{m_J}\right)\,\,\,(m_{J} \simeq {\cal O}(\unit{keV}))$ \\
                  &              $g_{\nu J} \lesssim 5 \times 10^{-5}\,\,\,(m_J \simeq {\cal O}(\unit{eV}))$ \cite{Huang:2017egl,Aghanim:2018eyx} \\
\hline\hline
{\bf Model B}: $i g_{eJ} \, \bar e \, \gamma_5 \, e \, J$ & \\
\hline
Solar lifetime & $g_{e J} \lesssim 3 \times 10^{-11}$~\cite{Gondolo:2008dd} \\
\hline
$\Delta N_{\rm eff}^{\rm BBN}$ & $g_{e J} \lesssim 5 \times 10^{-5}$ \cite{Huang:2017egl,Aghanim:2018eyx} \\
\hline
$(g-2)_e$ & $g_{eJ} \lesssim 1.8 \times 10^{-5}$~\cite{Parker191,Liu:2018xkx} \\
\hline\hline
{\bf Model C}: $g_{\nu L} \, \bar \nu \gamma^\mu P_L  \nu Z'_\mu$ & \\
\hline
$Z$ decay & $g_{\nu L} \lesssim 3 \times 10^{-2}$~\cite{Laha:2013xua} \\
\hline
$W$ decays & $g_{\nu L} \lesssim 2.5 \times 10^{-7}\left(\frac{m_{Z'}}{\unit[1]{keV}}\right)$~\cite{Laha:2013xua} \\
\hline
Meson decays & $g_{\nu L} \lesssim 6 \times 10^{-7} \left(\frac{m_{Z'}}{\unit[1]{keV}}\right)$~\cite{Bakhti:2017jhm,Bakhti:2018avv} \\
\hline
$\Delta N_{\rm eff}^{\rm BBN}$ & $g_{\nu L} \lesssim 2.2 \times 10^{-7} \left(\frac{\unit[1]{keV}}{m_{Z'}}\right)\,\,\left(m_{Z'}\simeq {\cal O}(\unit{keV})\right)$\\
  & $g_{\nu L} \lesssim 4.6 \times 10^{-6}\,\,\left(m_{Z'} \simeq {\cal O}(\unit{eV})\right)$~\cite{Huang:2017egl}\\
\hline
CMB &  $g_{\nu L} \lesssim 1.2 \times 10^{-2} \left(\frac{m_{Z'}}{\unit[1]{MeV}}\right)\,\,\left(m_{Z'} \gg \unit[1]{eV}\right)$~\cite{Archidiacono:2013dua}\\
\hline
Supernova 1987A & $g_{\nu L}\lesssim 12 \left(\frac{m_{Z'}}{\unit[1]{MeV}}\right)\,\,\left(m_{Z'}\geq \unit[60]{eV}\right)$\\
 & $g_{\nu L} \lesssim 5.6 \times 10^{-4}\,\,\left(m_{Z'} < \unit[60]{eV}\right)$~\cite{Kolb:1987qy}\\
\hline

IceCube & $g_{\nu L} \lesssim 0.03$~\cite{Ioka:2014kca}\\
\hline\hline
{\bf Model D}: $g_{eV} \bar e \gamma^\mu  e Z'_\mu$ & \\
\hline
$W$ decays & $g_{eV} \lesssim 2.5 \times 10^{-7}\left(\frac{m_{Z'}}{\unit[1]{keV}}\right)$~\cite{Laha:2013xua}\\
\hline
Solar lifetime & $g_{eV} \lesssim 2 \times 10^{-13}$~\cite{Gondolo:2008dd} \\
\hline
$\Delta N_{\rm eff}^{\rm BBN}$ & $g_{eV} \lesssim 4.6 \times 10^{-6}$~\cite{Huang:2017egl} \\
\hline
$(g-2)_e$ & $g_{eV} \lesssim 4.0 \times 10^{-6}$~\cite{Parker191,Liu:2018xkx}\\
\hline\hline
{\bf Model E}: $g_{L_e}\left(\bar \nu_e \gamma^\mu P_L  \nu_e+ \bar e \gamma^\mu e\right)Z'_\mu$ & \\
\hline
$Z$ decay & $g_{L_e} \lesssim 3 \times 10^{-2}$~\cite{Laha:2013xua} \\
\hline
$\Delta N_{\rm eff}^{\rm BBN,\,CMB}$ &  as models C and D\\
\hline
$\nu$--$e$ scattering & $g_{L_e} \lesssim 10^{-6}$~\cite{Laha:2013xua,Lindner:2018kjo} \\
\hline\hline
\end{tabular}
\caption{Summary table including all the laboratory, astroparticle, and cosmological constraints which apply to the models considered in this work; LNV $=$ lepton number violating coupling.}
\label{tab:constraints}
\end{table*}

The set of constraints discussed in this section is summarized in Tab.~\ref{tab:constraints}. Their impact on the sensitivity of the~\katrin\, experiment, on the various models, is shown in Figs.~\ref{fig:SensitivityPseudoscalar-eV} (\ref{fig:sensitivity_keV_pseudoscalar}) and \ref{fig:SensitivityVectorboson-eV} (\ref{fig:sensitivity_keV_vector}), respectively for the cases of scalar (models A and B) and vector (models C, D and E) new states with masses $\leq {\cal O}(\unit{eV})$ (${\cal O}(\unit{keV})$). 

As evident, for all the considered scenarios, the sensitivity region of the~\katrin\, experiment appears to be already excluded by other laboratory searches as well by cosmology and astrophysical observations. We notice, nevertheless, that most of the laboratory constraints here listed have been extrapolated from searches of heavier states. In addition, the energy scale of the processes was orders of magnitude larger than the one of tritium decay. 
In this regard, the~\katrin\, detector, which is designed for the search of very light states, would then provide a more solid and complementary constraint. Along a similar reasoning, it would provide a laboratory complement to the astrophysical and cosmological limits, which rely on specific hypotheses.

\section{\label{sec:concl}Conclusion}
\noindent In this paper we have shown that there is interesting physics potential in the \katrin~experiment 
beyond the neutrino mass hunt. Emission of light scalar or vector bosons from the neutrino or electron lines can modify the energy spectrum of the electrons in tritium decay and produce observable signals. We have calculated the spectra and performed a detailed analysis of the sensitivity of the \katrin~experiment for light particles around eV. A future \katrin~setup investigating the full electron spectrum was also investigated and the statistical sensitivity for keV-scale light particles obtained. 

The obtainable constraints are not competitive with high energy laboratory searches, like e.g.\ from decays of weak gauge bosons, as well as cosmological and astrophysical constraints. Nevertheless, they represent a solid complementary approach to rarely studied low energy new physics, performed at a scale that corresponds to the new physics scale, i.e.\ without the need of extrapolation.

\section*{Acknowledgements}\noindent
JH is a postdoctoral researcher of the F.R.S.-FNRS and furthermore supported, in part, by the National Science Foundation under Grant No.~PHY-1620638, and by a Feodor Lynen Research Fellowship of the Alexander von Humboldt Foundation. WR is supported by the DFG with grant RO~2516/7-1 in the Heisenberg program. FSQ acknowledges support from MEC, UFRN and ICTP-SAIFR FAPESP grant 2016/01343-7. KV and FH are supported by the Helmholtz Young Investigator Group VH-NG-1055. MS and FH are grateful to W.Q.~Choi and A.~Lokhov for very valuable discussions.

\appendix

\section{Formulae for the spectrum}
\label{sec:appendix}
\noindent
The decay of a particle $\mathcal{A}$ with momentum $p_\mathcal{A}$ into four particles with momenta $p_{1,2,3,4}$ is given by an amplitude $\mathcal{M}$, from which we can calculate the spin-averaged amplitude-squared $|\mathcal{M}|^2$ as a function of all momenta. The differential decay rate is then given by
\begin{align}
 \dd \Gamma = \frac{1}{2 m_\mathcal{A}} \frac{|\mathcal{M}|^2}{(2\pi)^8} \, \delta^{(4)} \left(p_\mathcal{A} -\sum_j p_j\right) \prod_j \frac{\dd^3 p_j}{2 E_j}\,.
\end{align}
An explicit parametrization of the four-body phase space was given long ago in Ref.~\cite{Nyborg:1965zz}, for which it is convenient to introduce the invariant masses
\begin{align}
	M_{i\dots j}^2 \equiv (E_i+\dots + E_j)^2-(\vec{p}_i+\dots +\vec{p}_j)^2 \,.
	\label{eq:mass-squares}
\end{align}
Only five of these are linearly independent, and we will choose $M_{12}^2$, $M_{34}^2$, $M_{14}^2$, $M_{124}^2$, and $M_{134}^2$ as our variables. It is a straightforward exercise to invert Eq.~\eqref{eq:mass-squares} and express $|\mathcal{M}|^2$ in these new variables. Performing all other integrations this leads to
\begin{align}
\dd \Gamma = \frac{\pi^2}{16 m_\mathcal{A}^3} \frac{|\mathcal{M}|^2}{(2\pi)^8}  \frac{1}{\sqrt{-B}} \dd M_{12}^2 \dd  M_{34}^2 \dd  M_{14}^2 \dd  M_{124}^2 \dd M_{134}^2 \,,
\end{align}
where $B$ is a lengthy negative function of all masses and mass-squares, with $B=0$ defining the boundary of the physically allowed region~\cite{Nyborg:1965zz}. Since $B$ is only quadratic in all the variables, it is typically possible to perform several of the integrals analytically. As we are interested in the electron spectrum, it behooves us to assign $p_e = p_3$ and change variables to $E_e$ using
\begin{align}
\hspace{-2.3mm}	M_{124}^2= m_\mathcal{A}^2+m_e^2-2 m_\mathcal{A} E_e \,, \ 
	\dd M_{124}^2= -2 m_\mathcal{A} \dd E_e \,.
\end{align}
All other momenta should be assigned in such a way that $|\mathcal{M}|^2$ becomes as simple as possible, ideally independent of some of the mass-squares, so that one can perform some of the remaining integrals analytically.

As an example, let us consider the case of tritium decay $\mathcal{A}\to \mathcal{B} +e^-+\overline{\nu}_e$, as governed by the Lagrangian
\begin{align}
	\L = -\frac{G_F V_{ud}}{\sqrt{2}} \left(\overline{e} \gamma^\mu (1-\gamma_5) \nu_e \right)\left(\overline{\mathcal{B}} \gamma_\mu (g_V-g_A\gamma_5) \mathcal{A} \right)+\hc	
\end{align}
 with emission of a pseudoscalar $J$ off the neutrino via a coupling $i g_{\nu J} \, \bar \nu_e  \gamma_5  \nu_e \, J$. Here, $\mathcal{A}={}^3\text{H}^{+}$ and $\mathcal{B}={}^3\text{He}^{2+}$ are treated as elementary fermions, see Ref.~\cite{Ludl:2016ane}. Assigning the momenta $p_1$, $p_2$, $p_3$, and $p_4$ to neutrino, pseudoscalar, electron, and $\mathcal{B}$, respectively, the amplitude takes the form
 \begin{align}
 \begin{split}
 	\mathcal{M} &= \frac{G_F V_{ud} g_{\nu J}}{\sqrt{2}} \overline{u}(p_4) \gamma_\mu (g_V-g_A\gamma_5) u(p_\mathcal{A})\\
 	& \quad \times \overline{u}(p_3) \gamma^\mu (1-\gamma_5) \frac{\slashed{p}_1+\slashed{p}_2}{(p_1+p_2)^2} \gamma_5 v(p_1) \,.
 	\end{split}
 \end{align}
The spin-averaged $|\mathcal{M}|^2$ is then linear in $M_{14}^2$ and can be integrated without much effort. The $M_{134}^2$ and $M_{34}^2$ integrals are trivial, leading to the final expression for the differential decay rate 
\begin{align}
\frac{\dd \Gamma}{\dd E_e} = \int_{m_J^2}^{\left(\sqrt{ m_\mathcal{A}^2-2 m_\mathcal{A} E_e+m_e^2} - m_\mathcal{B}\right)^2} I \,\dd M_{12}^2  
\label{eq:differential_rate_Jnu}
\end{align}
with rather lengthy integrand
\begin{widetext}
\begin{align}
	I &= \frac{G_F^2 g_{\nu J}^2 |V_{ud}|^2 \sqrt{E_e^2-m_e^2} \left(M_{12}^2-m_J^2\right)^2}{3072 \pi ^5 (M_{12}^2)^3 m_\mathcal{A} \left(m_\mathcal{A}^2-2 E_e m_\mathcal{A}+m_e^2\right)^3} \left[ g_V^2 m_\mathcal{A} W_V+2 g_V g_A W_{VA}+g_A^2 m_\mathcal{A} W_A \right] \\ \nonumber
	&\quad \times \sqrt{-2 m_\mathcal{B}^2 \left(-2 E_e m_\mathcal{A}+M_{12}^2+m_\mathcal{A}^2+m_e^2\right)+\left(-2 E_e m_\mathcal{A}-M_{12}^2+m_\mathcal{A}^2+m_e^2\right)^2+m_\mathcal{B}^4} \,,\\
	W_V &= -6 m_\mathcal{A}^2 m_\mathcal{B} \left(-2 E_e m_\mathcal{A}+m_\mathcal{A}^2+m_e^2\right) \left(-2 E_e m_\mathcal{A}+M_{12}^2+m_\mathcal{A}^2-m_\mathcal{B}^2+m_e^2\right)\nonumber\\\nonumber
	&\quad +m_\mathcal{A} \left[-2 \left(-2 E_e m_\mathcal{A}+m_\mathcal{A}^2+m_e^2\right)^2 \left(4 \left(M_{12}^2+m_\mathcal{B}^2\right)+m_e^2\right)+4 m_e^2 \left(M_{12}^2-m_\mathcal{B}^2\right)^2\right.\\\nonumber
	&\quad \left.+\left(-2 E_e m_\mathcal{A}+m_\mathcal{A}^2+m_e^2\right) \left(\left(M_{12}^2-m_\mathcal{B}^2\right)^2-2 m_e^2 \left(M_{12}^2+m_\mathcal{B}^2\right)\right)+7 \left(-2 E_e m_\mathcal{A}+m_\mathcal{A}^2+m_e^2\right)^3\right]\\\nonumber
	&\quad +6 m_\mathcal{B} \left(-2 E_e m_\mathcal{A}+m_\mathcal{A}^2+m_e^2\right) \left(-2 E_e m_\mathcal{A}+m_\mathcal{A}^2+2 m_e^2\right) \left(-2 E_e m_\mathcal{A}+M_{12}^2+m_\mathcal{A}^2-m_\mathcal{B}^2+m_e^2\right)\\
	&\quad +(2 E_e-m_\mathcal{A}) \left[\left(-2 E_e m_\mathcal{A}+m_\mathcal{A}^2+m_e^2\right)^2 \left(m_e^2-7 \left(M_{12}^2+m_\mathcal{B}^2\right)\right)\right.\\\nonumber
	&\quad +\left(-2 E_e m_\mathcal{A}+m_\mathcal{A}^2+m_e^2\right) \left(m_e^2 \left(M_{12}^2+m_\mathcal{B}^2\right)-\left(M_{12}^2-m_\mathcal{B}^2\right)^2\right)\\\nonumber
	&\quad\left.+8 \left(-2 E_e m_\mathcal{A}+m_\mathcal{A}^2+m_e^2\right)^3-2 m_e^2 \left(M_{12}^2-m_\mathcal{B}^2\right)^2\right]\\\nonumber
	&\quad +m_\mathcal{A}^3 \left(\left(M_{12}^2+m_\mathcal{B}^2\right) \left(-2 E_e m_\mathcal{A}+m_\mathcal{A}^2+m_e^2\right)+\left(-2 E_e m_\mathcal{A}+m_\mathcal{A}^2+m_e^2\right)^2-2 \left(M_{12}^2-m_\mathcal{B}^2\right)^2\right) ,\\\nonumber
	W_{VA} &= -2 \left(M_{12}^2-m_\mathcal{B}^2\right)^2 \left(m_\mathcal{A}^2-m_e^2\right)^2 -5 \left(-2 E_e m_\mathcal{A}+m_\mathcal{A}^2+m_e^2\right)^3 \left(M_{12}^2+m_\mathcal{A}^2+m_\mathcal{B}^2+m_e^2\right)\\\nonumber
	&\quad +4 \left(-2 E_e m_\mathcal{A}+m_\mathcal{A}^2+m_e^2\right)^4+\left(-2 E_e m_\mathcal{A}+m_\mathcal{A}^2+m_e^2\right)^2 \left[(M_{12}^2)^2+2 m_\mathcal{A}^2 \left(2 \left(M_{12}^2+m_\mathcal{B}^2\right)-m_e^2\right)\right.\\ 
	&\quad\left.-2 M_{12}^2 m_\mathcal{B}^2+4 M_{12}^2 m_e^2+m_\mathcal{A}^4+m_\mathcal{B}^4+4 m_\mathcal{B}^2 m_e^2+m_e^4\right]+\left(-2 E_e m_\mathcal{A}+m_\mathcal{A}^2+m_e^2\right) \left[m_\mathcal{A}^4 \left(M_{12}^2+m_\mathcal{B}^2\right)\right.\\ \nonumber
	&\quad\left.+m_\mathcal{A}^2 \left(\left(M_{12}^2-m_\mathcal{B}^2\right)^2-2 m_e^2 \left(M_{12}^2+m_\mathcal{B}^2\right)\right)+m_e^4 \left(M_{12}^2+m_\mathcal{B}^2\right)+m_e^2 \left(M_{12}^2-m_\mathcal{B}^2\right)^2\right],\\ \nonumber 
	W_A &= W_V \text{ with } m_\mathcal{B}\to -m_\mathcal{B} \,.
\end{align}
\end{widetext}
The remaining integral over $M_{12}^2$ can be performed numerically to obtain $\dd \Gamma/\dd E_e$. 

At this point one can implement a correction to account for the electromagnetic interaction of the emitted electron with the newly formed ${}^3\text{He}^{2+}$ nucleus,
\begin{align}
	\frac{\dd \Gamma}{\dd E_e} \to \frac{\dd \Gamma}{\dd E_e} F(Z,E_e)\,,
\end{align}
with the Fermi function~\cite{Ludl:2016ane}
\begin{align}
	F(Z,E_e) = 2 (1+\gamma)\frac{e^{\pi y}}{(2 p_e R)^{2 (1-\gamma)}} \frac{|\Gamma (\gamma+i y)|^2}{\Gamma (2\gamma+1)^2}\,,
\label{eq:fermi}
\end{align}
with coefficients $y= Z \alpha E_e/p_e$ and $\gamma =(1- Z^2\alpha^2)^{1/2}$ as well as the Gamma function $\Gamma$, not to be confused with the decay rate. The radius and electric charge of the  ${}^3\text{He}^{2+}$ nucleus are given by $R\simeq 2.884\times 10^{-3}/m_e$ and $Z=2$, respectively. For the standard tritium decay this correction improves the accuracy to the percent level~\cite{Ludl:2016ane}, certainly sufficient for our purposes.

With the so-obtained differential distribution the total decay rate can finally be obtained via
\begin{align}
\Gamma (\mathcal{A}\to \mathcal{B} +e^-+\overline{\nu}_e + J)  = \int_{m_e}^{E_e^\text{max}} \frac{\dd \Gamma}{\dd E_e} \dd E_e 
\end{align}
with endpoint energy of Eq.~\eqref{eq:endpoint_energy}.
With the above expressions we can numerically integrate $\dd \Gamma/\dd E_e$ to very good precision, despite the small available phase space in tritium decay.

Emitting a gauge boson $Z'$ from the neutrino via $g_{\nu L}  \bar \nu \gamma^\mu P_L  \nu Z'_\mu $ instead of a pseudoscalar results in a similar distribution as Eq.~\eqref{eq:differential_rate_Jnu} and can be obtained from the former via
\begin{align}
	\frac{\dd \Gamma}{\dd E_e}\bigg\rvert_{Z'} = \frac{g_{\nu L}^2 \left(M_{12}^2 +2 m_J^2\right)}{g_{\nu J}^2 m_J^2}
	\frac{\dd \Gamma}{\dd E_e}\bigg\rvert_{J}
\end{align}
and replacing $m_J\to m_{Z'}$ everywhere. One consequence of this relation is a different behavior in the limit of ultra-light bosons: for the pseudoscalar, $\Gamma (\mathcal{A}\to \mathcal{B} +e^-+\overline{\nu}_e + J) $ is roughly constant for $m_J\to 0$, except for a small logarithmic collinear divergence. In the gauge boson case, however, $\Gamma (\mathcal{A}\to \mathcal{B} +e^-+\overline{\nu}_e + Z')$ grows with $1/m_{Z'}^2$ for small $Z'$ mass. This is of course a well known behavior of gauge boson couplings to a non-conserved current~\cite{Dror:2017ehi,Dror:2017nsg}, as is the case here. Indeed, if the $Z'$ couples to the classically \emph{conserved} electron-number current $j^\alpha_{L_e}=\bar \nu_e \gamma^\alpha P_L  \nu_e+ \bar e \gamma^\alpha e$ then $\Gamma (\mathcal{A}\to \mathcal{B} +e^-+\overline{\nu}_e + Z')$ remains constant for small $Z'$ mass, up to small logarithmic corrections.

Tritium decay with additional boson emission off the neutrino is the simplest process to calculate, but from the above it is evident that the expressions are still unwieldy when all masses are kept non-zero (except the neutrino mass). We will therefore not give analytical expressions for the more complicated case of boson emission off the electron. 
Luckily, all spectra of interest can be described to excellent precision by the simple function
\begin{align}
\frac{\dd \Gamma}{\dd E_e} = \frac{K}{\hbar} \sqrt{\frac{E_e}{m_e}-1} \, \left(1- \frac{E_e}{E_e^\text{max}}\right)^n F(Z,E_e)\,,
\label{eq:fit_function}
\end{align}
reintroducing the reduced Planck constant $\hbar$ for convenience.
Here, $K$ is a dimensionless normalization prefactor and $n$ the shape or spectral index. Both parameters depend on the new boson mass and, of course, on the model, but can be readily fitted to our numerically obtained spectra. We show the results in Fig.~\ref{fig:fit_parameters}; the spectral index $n$ lies between 2 and 4.5 for all our models, which implies rather similar looking spectra. The main difference of the models is indeed the normalization, as is evident already from Fig.~\ref{fig:all_branching_ratios}.
For comparison, the SM beta decay can be fitted rather well with $n\simeq 2$ and $K\simeq 1.26\times 10^{-24}$.

\bibliographystyle{utcaps_mod}
\bibliography{BIB}

\providecommand{\href}[2]{#2}\begingroup\raggedright\begin{thebibliography}{10}

\bibitem{Drexlin:2013lha}
G.~Drexlin, V.~Hannen, S.~Mertens, and C.~Weinheimer, ``{\em {Current direct
  neutrino mass experiments}},''
  \href{http://dx.doi.org/10.1155/2013/293986}{Adv. High Energy Phys.
  {\normalfont \bfseries 2013} (2013)  293986},
\href{http://arxiv.org/abs/1307.0101}{{\normalfont \ttfamily arXiv:1307.0101}}.

\bibitem{Osipowicz:2001sq}
{\normalfont \bfseries KATRIN}, A.~Osipowicz {\em et al.}, ``{\em {KATRIN: A
  Next generation tritium beta decay experiment with sub-eV sensitivity for the
  electron neutrino mass. Letter of intent}},''
\href{http://arxiv.org/abs/hep-ex/0109033}{{\normalfont \ttfamily
  arXiv:hep-ex/0109033}}.

\bibitem{Angrik:2005ep}
{\normalfont \bfseries KATRIN}, J.~Angrik {\em et al.}, ``{\em {KATRIN design
  report 2004}},''.
\url{https://publikationen.bibliothek.kit.edu/270060419/3814644}.

\bibitem{Riis:2010zm}
A.~S. Riis and S.~Hannestad, ``{\em {Detecting sterile neutrinos with KATRIN
  like experiments}},''
  \href{http://dx.doi.org/10.1088/1475-7516/2011/02/011}{JCAP {\normalfont
  \bfseries 1102} (2011)  011},
\href{http://arxiv.org/abs/1008.1495}{{\normalfont \ttfamily arXiv:1008.1495}}.

\bibitem{Formaggio:2011jg}
J.~A. Formaggio and J.~Barrett, ``{\em {Resolving the Reactor Neutrino Anomaly
  with the KATRIN Neutrino Experiment}},''
  \href{http://dx.doi.org/10.1016/j.physletb.2011.10.069}{Phys. Lett.
  {\normalfont \bfseries B706} (2011)  68--71},
\href{http://arxiv.org/abs/1105.1326}{{\normalfont \ttfamily arXiv:1105.1326}}.

\bibitem{SejersenRiis:2011sj}
A.~Sejersen~Riis, S.~Hannestad, and C.~Weinheimer, ``{\em {Analysis of
  simulated data for the KArlsruhe TRItium Neutrino experiment using Bayesian
  inference}},'' \href{http://dx.doi.org/10.1103/PhysRevC.84.045503}{Phys. Rev.
  {\normalfont \bfseries C84} (2011)  045503},
\href{http://arxiv.org/abs/1105.6005}{{\normalfont \ttfamily arXiv:1105.6005}}.

\bibitem{Esmaili:2012vg}
A.~Esmaili and O.~L.~G. Peres, ``{\em {KATRIN Sensitivity to Sterile Neutrino
  Mass in the Shadow of Lightest Neutrino Mass}},''
  \href{http://dx.doi.org/10.1103/PhysRevD.85.117301}{Phys. Rev. {\normalfont
  \bfseries D85} (2012)  117301},
\href{http://arxiv.org/abs/1203.2632}{{\normalfont \ttfamily arXiv:1203.2632}}.

\bibitem{Diaz:2014hca}
J.~S. Díaz, ``{\em {Tests of Lorentz symmetry in single beta decay}},''
  \href{http://dx.doi.org/10.1155/2014/305298}{Adv. High Energy Phys.
  {\normalfont \bfseries 2014} (2014)  305298},
\href{http://arxiv.org/abs/1408.5880}{{\normalfont \ttfamily arXiv:1408.5880}}.

\bibitem{Rodejohann:2014eka}
W.~Rodejohann and H.~Zhang, ``{\em {Signatures of Extra Dimensional Sterile
  Neutrinos}},'' \href{http://dx.doi.org/10.1016/j.physletb.2014.08.035}{Phys.
  Lett. {\normalfont \bfseries B737} (2014)  81--89},
\href{http://arxiv.org/abs/1407.2739}{{\normalfont \ttfamily arXiv:1407.2739}}.

\bibitem{Steinbrink:2017uhw}
N.~M.~N. Steinbrink, F.~Glück, F.~Heizmann, M.~Kleesiek, K.~Valerius,
  C.~Weinheimer, and S.~Hannestad, ``{\em {Statistical sensitivity on
  right-handed currents in presence of eV scale sterile neutrinos with
  KATRIN}},'' \href{http://dx.doi.org/10.1088/1475-7516/2017/06/015}{JCAP
  {\normalfont \bfseries 1706} (2017)  015},
\href{http://arxiv.org/abs/1703.07667}{{\normalfont \ttfamily
  arXiv:1703.07667}}.

\bibitem{Mertens:2014nha}
S.~Mertens, T.~Lasserre, S.~Groh, G.~Drexlin, F.~Glueck, A.~Huber, A.~W.~P.
  Poon, M.~Steidl, N.~Steinbrink, and C.~Weinheimer, ``{\em {Sensitivity of
  Next-Generation Tritium Beta-Decay Experiments for keV-Scale Sterile
  Neutrinos}},'' \href{http://dx.doi.org/10.1088/1475-7516/2015/02/020}{JCAP
  {\normalfont \bfseries 1502} (2015)  020},
\href{http://arxiv.org/abs/1409.0920}{{\normalfont \ttfamily arXiv:1409.0920}}.

\bibitem{Mertens:2018}
S.~Mertens {\em et al.}, ``{\em A novel detector system for KATRIN to search
  for keV-scale sterile neutrinos},''
  \href{http://arxiv.org/abs/1810.06711}{{\normalfont \ttfamily
  arXiv:1810.06711}}.

\bibitem{Adhikari:2016bei}
M.~Drewes {\em et al.}, ``{\em {A White Paper on keV Sterile Neutrino Dark
  Matter}},'' \href{http://dx.doi.org/10.1088/1475-7516/2017/01/025}{JCAP
  {\normalfont \bfseries 1701} (2017)  025},
\href{http://arxiv.org/abs/1602.04816}{{\normalfont \ttfamily
  arXiv:1602.04816}}.

\bibitem{Shrock:1980vy}
R.~E. Shrock, ``{\em {New Tests For, and Bounds On, Neutrino Masses and Lepton
  Mixing}},''
\href{http://dx.doi.org/10.1016/0370-2693(80)90235-X}{Phys. Lett. {\normalfont
  \bfseries 96B} (1980)  159--164}.

\bibitem{Herczeg:2001vk}
P.~Herczeg, ``{\em {Beta decay beyond the standard model}},''
\href{http://dx.doi.org/10.1016/S0146-6410(01)00149-1}{Prog. Part. Nucl. Phys.
  {\normalfont \bfseries 46} (2001)  413--457}.

\bibitem{Severijns:2006dr}
N.~Severijns, M.~Beck, and O.~Naviliat-Cuncic, ``{\em {Tests of the standard
  electroweak model in beta decay}},''
  \href{http://dx.doi.org/10.1103/RevModPhys.78.991}{Rev. Mod. Phys.
  {\normalfont \bfseries 78} (2006)  991--1040},
\href{http://arxiv.org/abs/nucl-ex/0605029}{{\normalfont \ttfamily
  arXiv:nucl-ex/0605029}}.

\bibitem{Liao:2010yx}
W.~Liao, ``{\em {keV scale $\nu_{R}$ dark matter and its detection in $\beta$
  decay experiment}},''
  \href{http://dx.doi.org/10.1103/PhysRevD.82.073001}{Phys. Rev. {\normalfont
  \bfseries D82} (2010)  073001},
\href{http://arxiv.org/abs/1005.3351}{{\normalfont \ttfamily arXiv:1005.3351}}.

\bibitem{deVega:2011xh}
H.~J. de~Vega, O.~Moreno, E.~M. de~Guerra, M.~R. Medrano, and N.~G. Sanchez,
  ``{\em {Role of sterile neutrino warm dark matter in rhenium and tritium beta
  decays}},'' \href{http://dx.doi.org/10.1016/j.nuclphysb.2012.08.019}{Nucl.
  Phys. {\normalfont \bfseries B866} (2013)  177--195},
\href{http://arxiv.org/abs/1109.3452}{{\normalfont \ttfamily arXiv:1109.3452}}.

\bibitem{Abdurashitov:2014vqa}
D.~N. Abdurashitov, A.~I. Berlev, N.~A. Likhovid, A.~V. Lokhov, I.~I. Tkachev,
  and V.~E. Yants, ``{\em {Searches for a Sterile-Neutrino Admixture in
  Detecting Tritium Decays in a Proportional Counter: New Possibilities}},''
  \href{http://dx.doi.org/10.1134/S1063778815020027}{Phys. Atom. Nucl.
  {\normalfont \bfseries 78} (2015) no.~2, 268--280},
\href{http://arxiv.org/abs/1403.2935}{{\normalfont \ttfamily arXiv:1403.2935}}.

\bibitem{Barry:2014ika}
J.~Barry, J.~Heeck, and W.~Rodejohann, ``{\em {Sterile neutrinos and
  right-handed currents in KATRIN}},''
  \href{http://dx.doi.org/10.1007/JHEP07(2014)081}{JHEP {\normalfont \bfseries
  07} (2014)  081},
\href{http://arxiv.org/abs/1404.5955}{{\normalfont \ttfamily arXiv:1404.5955}}.

\bibitem{Gonzalez-Alonso:2018omy}
M.~Gonzalez-Alonso, O.~Naviliat-Cuncic, and N.~Severijns, ``{\em {New physics
  searches in nuclear and neutron $\beta$ decay}},''
\href{http://arxiv.org/abs/1803.08732}{{\normalfont \ttfamily
  arXiv:1803.08732}}.

\bibitem{Abada:2018qok}
A.~Abada, A.~Hernández-Cabezudo, and X.~Marcano, ``{\em {Beta and Neutrinoless
  Double Beta Decays with KeV Sterile Fermions}},''
\href{http://arxiv.org/abs/1807.01331}{{\normalfont \ttfamily
  arXiv:1807.01331}}.

\bibitem{Ludl:2016ane}
P.~O. Ludl and W.~Rodejohann, ``{\em {Direct Neutrino Mass Experiments and
  Exotic Charged Current Interactions}},''
  \href{http://dx.doi.org/10.1007/JHEP06(2016)040}{JHEP {\normalfont \bfseries
  06} (2016)  040},
\href{http://arxiv.org/abs/1603.08690}{{\normalfont \ttfamily
  arXiv:1603.08690}}.

\bibitem{Masood:2007rc}
S.~S. Masood, S.~Nasri, J.~Schechter, M.~A. Tortola, J.~W.~F. Valle, and
  C.~Weinheimer, ``{\em {Exact relativistic beta decay endpoint spectrum}},''
  \href{http://dx.doi.org/10.1103/PhysRevC.76.045501}{Phys. Rev. {\normalfont
  \bfseries C76} (2007)  045501},
\href{http://arxiv.org/abs/0706.0897}{{\normalfont \ttfamily arXiv:0706.0897}}.

\bibitem{Chikashige:1980ui}
Y.~Chikashige, R.~N. Mohapatra, and R.~D. Peccei, ``{\em {Are There Real
  Goldstone Bosons Associated with Broken Lepton Number?}},''
\href{http://dx.doi.org/10.1016/0370-2693(81)90011-3}{Phys. Lett. {\normalfont
  \bfseries 98B} (1981)  265--268}.

\bibitem{Schechter:1981cv}
J.~Schechter and J.~W.~F. Valle, ``{\em {Neutrino Decay and Spontaneous
  Violation of Lepton Number}},''
\href{http://dx.doi.org/10.1103/PhysRevD.25.774}{Phys. Rev. {\normalfont
  \bfseries D25} (1982)  774}.

\bibitem{Pilaftsis:1993af}
A.~Pilaftsis, ``{\em {Astrophysical and terrestrial constraints on singlet
  Majoron models}},'' \href{http://dx.doi.org/10.1103/PhysRevD.49.2398}{Phys.
  Rev. {\normalfont \bfseries D49} (1994)  2398--2404},
\href{http://arxiv.org/abs/hep-ph/9308258}{{\normalfont \ttfamily
  arXiv:hep-ph/9308258}}.

\bibitem{Garcia-Cely:2017oco}
C.~Garcia-Cely and J.~Heeck, ``{\em {Neutrino Lines from Majoron Dark
  Matter}},'' \href{http://dx.doi.org/10.1007/JHEP05(2017)102}{JHEP
  {\normalfont \bfseries 05} (2017)  102},
\href{http://arxiv.org/abs/1701.07209}{{\normalfont \ttfamily
  arXiv:1701.07209}}.

\bibitem{Langacker:2008yv}
P.~Langacker, ``{\em {The Physics of Heavy $Z^\prime$ Gauge Bosons}},''
  \href{http://dx.doi.org/10.1103/RevModPhys.81.1199}{Rev. Mod. Phys.
  {\normalfont \bfseries 81} (2009)  1199--1228},
\href{http://arxiv.org/abs/0801.1345}{{\normalfont \ttfamily arXiv:0801.1345}}.

\bibitem{Ruegg:2003ps}
H.~Ruegg and M.~Ruiz-Altaba, ``{\em {The Stueckelberg field}},''
  \href{http://dx.doi.org/10.1142/S0217751X04019755}{Int. J. Mod. Phys.
  {\normalfont \bfseries A19} (2004)  3265--3348},
\href{http://arxiv.org/abs/hep-th/0304245}{{\normalfont \ttfamily
  arXiv:hep-th/0304245}}.

\bibitem{Alexander:2016aln}
J.~Alexander {\em et al.}, ``{\em {Dark Sectors 2016 Workshop: Community
  Report}},''
\newblock 2016.
\newblock
\href{http://arxiv.org/abs/1608.08632}{{\normalfont \ttfamily
  arXiv:1608.08632}}.
\newblock

\bibitem{Farzan:2016wym}
Y.~Farzan and J.~Heeck, ``{\em {Neutrinophilic nonstandard interactions}},''
  \href{http://dx.doi.org/10.1103/PhysRevD.94.053010}{Phys. Rev. {\normalfont
  \bfseries D94} (2016)  053010},
\href{http://arxiv.org/abs/1607.07616}{{\normalfont \ttfamily
  arXiv:1607.07616}}.

\bibitem{Bakhti:2018avv}
P.~Bakhti, Y.~Farzan, and M.~Rajaee, ``{\em {Secret interactions of neutrinos
  with light gauge boson at the DUNE near detector}},''
\href{http://arxiv.org/abs/1810.04441}{{\normalfont \ttfamily
  arXiv:1810.04441}}.

\bibitem{Dror:2017ehi}
J.~A. Dror, R.~Lasenby, and M.~Pospelov, ``{\em {New constraints on light
  vectors coupled to anomalous currents}},''
  \href{http://dx.doi.org/10.1103/PhysRevLett.119.141803}{Phys. Rev. Lett.
  {\normalfont \bfseries 119} (2017)  141803},
\href{http://arxiv.org/abs/1705.06726}{{\normalfont \ttfamily
  arXiv:1705.06726}}.

\bibitem{Dror:2017nsg}
J.~A. Dror, R.~Lasenby, and M.~Pospelov, ``{\em {Dark forces coupled to
  nonconserved currents}},''
  \href{http://dx.doi.org/10.1103/PhysRevD.96.075036}{Phys. Rev. {\normalfont
  \bfseries D96} (2017)  075036},
\href{http://arxiv.org/abs/1707.01503}{{\normalfont \ttfamily
  arXiv:1707.01503}}.

\bibitem{Kleesiek:2018mel}
M.~Kleesiek {\em et al.}, ``{\em {$\beta$-Decay Spectrum, Response Function and
  Statistical Model for Neutrino Mass Measurements with the KATRIN
  Experiment}},''
\href{http://arxiv.org/abs/1806.00369}{{\normalfont \ttfamily
  arXiv:1806.00369}}.

\bibitem{Repko:1984cs}
W.~W. Repko and C.-E. Wu, ``{\em {Radiative Corrections To The Endpoint Of The
  Tritium Beta Decay Spectrum}},''
\href{http://dx.doi.org/10.1103/PhysRevC.28.2433}{Phys. Rev. {\normalfont
  \bfseries C28} (1983)  2433--2436}.

\bibitem{Saenz:2000dul}
A.~Saenz, S.~Jonsell, and P.~Froelich, ``{\em {Improved Molecular Final-State
  Distribution of HeT$^+$ for the $\beta$-Decay Process of $T_2$}},''
\href{http://dx.doi.org/10.1103/PhysRevLett.84.242}{Phys. Rev. Lett.
  {\normalfont \bfseries 84} (2000)  242}.

\bibitem{Doss:2006zv}
N.~Doss, J.~Tennyson, A.~Saenz, and S.~Jonsell, ``{\em {Molecular effects in
  investigations of tritium molecule beta decay endpoint experiments}},''
\href{http://dx.doi.org/10.1103/PhysRevC.73.025502}{Phys. Rev. {\normalfont
  \bfseries C73} (2006)  025502}.

\bibitem{Doss2008}
N.~Doss and J.~Tennyson, ``{\em Excitations to the electronic continuum of
  $\mathrm{^3HeT^+}$ in investigations of $\mathrm{T}_2$ \textbeta{}-decay
  experiments},'' \href{http://dx.doi.org/10.1088/0953-4075/41/12/125701}{J.
  Phys. B {\normalfont \bfseries 41} (2008) no.~12, 125701+}.

\bibitem{Rolke:2004mj}
W.~A. Rolke, A.~M. Lopez, and J.~Conrad, ``{\em {Limits and confidence
  intervals in the presence of nuisance parameters}},''
  \href{http://dx.doi.org/10.1016/j.nima.2005.05.068}{Nucl. Instrum. Meth.
  {\normalfont \bfseries A551} (2005)  493--503},
\href{http://arxiv.org/abs/physics/0403059}{{\normalfont \ttfamily
  arXiv:physics/0403059}}.

\bibitem{Harms:2015}
F.~Harms, {\em Characterization and minimization of background processes in the
  KATRIN main spectrometer}.
\newblock PhD thesis, Karlsruhe Institute of Technology, 2015.

\bibitem{Laha:2013xua}
R.~Laha, B.~Dasgupta, and J.~F. Beacom, ``{\em {Constraints on New Neutrino
  Interactions via Light Abelian Vector Bosons}},''
  \href{http://dx.doi.org/10.1103/PhysRevD.89.093025}{Phys. Rev. {\normalfont
  \bfseries D89} (2014)  093025},
\href{http://arxiv.org/abs/1304.3460}{{\normalfont \ttfamily arXiv:1304.3460}}.

\bibitem{Bakhti:2017jhm}
P.~Bakhti and Y.~Farzan, ``{\em {Constraining secret gauge interactions of
  neutrinos by meson decays}},''
  \href{http://dx.doi.org/10.1103/PhysRevD.95.095008}{Phys. Rev. {\normalfont
  \bfseries D95} (2017)  095008},
\href{http://arxiv.org/abs/1702.04187}{{\normalfont \ttfamily
  arXiv:1702.04187}}.

\bibitem{Pasquini:2015fjv}
P.~S. Pasquini and O.~L.~G. Peres, ``{\em {Bounds on Neutrino-Scalar Yukawa
  Coupling}},'' \href{http://dx.doi.org/10.1103/PhysRevD.93.053007}{Phys. Rev.
  {\normalfont \bfseries D93} (2016)  053007},
  \href{http://arxiv.org/abs/1511.01811}{{\normalfont \ttfamily
  arXiv:1511.01811}}.
[Erratum: Phys. Rev.D93,079902(2016)].

\bibitem{Lessa:2007up}
A.~P. Lessa and O.~L.~G. Peres, ``{\em {Revising limits on neutrino-Majoron
  couplings}},'' \href{http://dx.doi.org/10.1103/PhysRevD.75.094001}{Phys. Rev.
  {\normalfont \bfseries D75} (2007)  094001},
\href{http://arxiv.org/abs/hep-ph/0701068}{{\normalfont \ttfamily
  arXiv:hep-ph/0701068}}.

\bibitem{Albert:2014fya}
{\normalfont \bfseries EXO-200}, J.~B. Albert {\em et al.}, ``{\em {Search for
  Majoron-emitting modes of double-beta decay of $^{136}$Xe with EXO-200}},''
  \href{http://dx.doi.org/10.1103/PhysRevD.90.092004}{Phys. Rev. {\normalfont
  \bfseries D90} (2014)  092004},
\href{http://arxiv.org/abs/1409.6829}{{\normalfont \ttfamily arXiv:1409.6829}}.

\bibitem{Tanabashi:2018oca}
{\normalfont \bfseries Particle Data Group}, M.~Tanabashi {\em et al.}, ``{\em
  {Review of Particle Physics}},''
\href{http://dx.doi.org/10.1103/PhysRevD.98.030001}{Phys. Rev. {\normalfont
  \bfseries D98} (2018)  030001}.

\bibitem{Parker191}
R.~H. Parker, C.~Yu, W.~Zhong, B.~Estey, and H.~M{\"u}ller, ``{\em Measurement
  of the fine-structure constant as a test of the Standard Model},''
  \href{http://dx.doi.org/10.1126/science.aap7706}{Science {\normalfont
  \bfseries 360} (2018) no.~6385, 191--195}.

\bibitem{Lindner:2016bgg}
M.~Lindner, M.~Platscher, and F.~S. Queiroz, ``{\em {A Call for New Physics:
  The Muon Anomalous Magnetic Moment and Lepton Flavor Violation}},''
  \href{http://dx.doi.org/10.1016/j.physrep.2017.12.001}{Phys. Rept.
  {\normalfont \bfseries 731} (2018)  1--82},
\href{http://arxiv.org/abs/1610.06587}{{\normalfont \ttfamily
  arXiv:1610.06587}}.

\bibitem{Liu:2018xkx}
J.~Liu, C.~E.~M. Wagner, and X.-P. Wang, ``{\em {A light complex scalar for the
  electron and muon anomalous magnetic moments}},''
\href{http://arxiv.org/abs/1810.11028}{{\normalfont \ttfamily
  arXiv:1810.11028}}.

\bibitem{Lindner:2018kjo}
M.~Lindner, F.~S. Queiroz, W.~Rodejohann, and X.-J. Xu, ``{\em
  {Neutrino-electron scattering: general constraints on $Z'$ and dark photon
  models}},'' \href{http://dx.doi.org/10.1007/JHEP05(2018)098}{JHEP
  {\normalfont \bfseries 05} (2018)  098},
\href{http://arxiv.org/abs/1803.00060}{{\normalfont \ttfamily
  arXiv:1803.00060}}.

\bibitem{Blum:2018ljv}
K.~Blum, Y.~Nir, and M.~Shavit, ``{\em {Neutrinoless double-beta decay with
  massive scalar emission}},''
  \href{http://dx.doi.org/10.1016/j.physletb.2018.08.022}{Phys. Lett.
  {\normalfont \bfseries B785} (2018)  354--361},
\href{http://arxiv.org/abs/1802.08019}{{\normalfont \ttfamily
  arXiv:1802.08019}}.

\bibitem{Brune:2018sab}
T.~Brune and H.~P{\"a}s, ``{\em {Majoron Dark Matter and Constraints on the
  Majoron-Neutrino Coupling}},''
\href{http://arxiv.org/abs/1808.08158}{{\normalfont \ttfamily
  arXiv:1808.08158}}.

\bibitem{Gondolo:2008dd}
P.~Gondolo and G.~Raffelt, ``{\em {Solar neutrino limit on axions and keV-mass
  bosons}},'' \href{http://dx.doi.org/10.1103/PhysRevD.79.107301}{Phys. Rev.
  {\normalfont \bfseries D79} (2009)  107301},
\href{http://arxiv.org/abs/0807.2926}{{\normalfont \ttfamily arXiv:0807.2926}}.

\bibitem{Redondo:2008aa}
J.~Redondo, ``{\em {Helioscope Bounds on Hidden Sector Photons}},''
  \href{http://dx.doi.org/10.1088/1475-7516/2008/07/008}{JCAP {\normalfont
  \bfseries 0807} (2008)  008},
\href{http://arxiv.org/abs/0801.1527}{{\normalfont \ttfamily arXiv:0801.1527}}.

\bibitem{An:2013yfc}
H.~An, M.~Pospelov, and J.~Pradler, ``{\em {New stellar constraints on dark
  photons}},'' \href{http://dx.doi.org/10.1016/j.physletb.2013.07.008}{Phys.
  Lett. {\normalfont \bfseries B725} (2013)  190--195},
\href{http://arxiv.org/abs/1302.3884}{{\normalfont \ttfamily arXiv:1302.3884}}.

\bibitem{Kolb:1987qy}
E.~W. Kolb and M.~S. Turner, ``{\em {Supernova SN 1987a and the Secret
  Interactions of Neutrinos}},''
\href{http://dx.doi.org/10.1103/PhysRevD.36.2895}{Phys. Rev. {\normalfont
  \bfseries D36} (1987)  2895}.

\bibitem{Ng:2014pca}
K.~C.~Y. Ng and J.~F. Beacom, ``{\em {Cosmic neutrino cascades from secret
  neutrino interactions}},''
  \href{http://dx.doi.org/10.1103/PhysRevD.90.065035}{Phys. Rev. {\normalfont
  \bfseries D90} (2014)  065035},
  \href{http://arxiv.org/abs/1404.2288}{{\normalfont \ttfamily
  arXiv:1404.2288}}.
[Erratum: Phys. Rev.D90,089904(2014)].

\bibitem{Ioka:2014kca}
K.~Ioka and K.~Murase, ``{\em {IceCube PeV–EeV neutrinos and secret
  interactions of neutrinos}},''
  \href{http://dx.doi.org/10.1093/ptep/ptu090}{PTEP {\normalfont \bfseries
  2014} (2014) no.~6, 061E01},
\href{http://arxiv.org/abs/1404.2279}{{\normalfont \ttfamily arXiv:1404.2279}}.

\bibitem{Gelmini:1982rr}
G.~B. Gelmini, S.~Nussinov, and M.~Roncadelli, ``{\em {Bounds and Prospects for
  the Majoron Model of Left-handed Neutrino Masses}},''
\href{http://dx.doi.org/10.1016/0550-3213(82)90107-9}{Nucl. Phys. {\normalfont
  \bfseries B209} (1982)  157--173}.

\bibitem{Choi:1987sd}
K.~Choi, C.~W. Kim, J.~Kim, and W.~P. Lam, ``{\em {Constraints on the Majoron
  Interactions From the Supernova {SN1987A}}},''
\href{http://dx.doi.org/10.1103/PhysRevD.37.3225}{Phys. Rev. {\normalfont
  \bfseries D37} (1988)  3225}.

\bibitem{Berezhiani:1989za}
Z.~G. Berezhiani and A.~{\relax Yu}. Smirnov, ``{\em {Matter Induced Neutrino
  Decay and Supernova {SN1987A}}},''
\href{http://dx.doi.org/10.1016/0370-2693(89)90052-X}{Phys. Lett. {\normalfont
  \bfseries B220} (1989)  279--284}.

\bibitem{Choi:1989hi}
K.~Choi and A.~Santamaria, ``{\em {Majorons and Supernova Cooling}},''
\href{http://dx.doi.org/10.1103/PhysRevD.42.293}{Phys. Rev. {\normalfont
  \bfseries D42} (1990)  293--306}.

\bibitem{Chang:1993yp}
S.~Chang and K.~Choi, ``{\em {Constraints from nucleosynthesis and SN1987A on
  majoron emitting double beta decay}},''
  \href{http://dx.doi.org/10.1103/PhysRevD.49.R12}{Phys. Rev. {\normalfont
  \bfseries D49} (1994)  12--15},
\href{http://arxiv.org/abs/hep-ph/9303243}{{\normalfont \ttfamily
  arXiv:hep-ph/9303243}}.

\bibitem{Kachelriess:2000qc}
M.~Kachelriess, R.~Tom{\`a}s, and J.~W.~F. Valle, ``{\em {Supernova bounds on
  Majoron emitting decays of light neutrinos}},''
  \href{http://dx.doi.org/10.1103/PhysRevD.62.023004}{Phys. Rev. {\normalfont
  \bfseries D62} (2000)  023004},
\href{http://arxiv.org/abs/hep-ph/0001039}{{\normalfont \ttfamily
  arXiv:hep-ph/0001039}}.

\bibitem{Tomas:2001dh}
R.~Tom{\`a}s, H.~P{\"a}s, and J.~W.~F. Valle, ``{\em {Generalized bounds on
  Majoron-neutrino couplings}},''
  \href{http://dx.doi.org/10.1103/PhysRevD.64.095005}{Phys. Rev. {\normalfont
  \bfseries D64} (2001)  095005},
\href{http://arxiv.org/abs/hep-ph/0103017}{{\normalfont \ttfamily
  arXiv:hep-ph/0103017}}.

\bibitem{Lindner:2001th}
M.~Lindner, T.~Ohlsson, and W.~Winter, ``{\em {Decays of supernova
  neutrinos}},'' \href{http://dx.doi.org/10.1016/S0550-3213(01)00603-4}{Nucl.
  Phys. {\normalfont \bfseries B622} (2002)  429--456},
\href{http://arxiv.org/abs/astro-ph/0105309}{{\normalfont \ttfamily
  arXiv:astro-ph/0105309}}.

\bibitem{Hannestad:2002ff}
S.~Hannestad, P.~Keranen, and F.~Sannino, ``{\em {A Supernova constraint on
  bulk Majorons}},'' \href{http://dx.doi.org/10.1103/PhysRevD.66.045002}{Phys.
  Rev. {\normalfont \bfseries D66} (2002)  045002},
\href{http://arxiv.org/abs/hep-ph/0204231}{{\normalfont \ttfamily
  arXiv:hep-ph/0204231}}.

\bibitem{Farzan:2002wx}
Y.~Farzan, ``{\em {Bounds on the coupling of the Majoron to light neutrinos
  from supernova cooling}},''
  \href{http://dx.doi.org/10.1103/PhysRevD.67.073015}{Phys. Rev. {\normalfont
  \bfseries D67} (2003)  073015},
\href{http://arxiv.org/abs/hep-ph/0211375}{{\normalfont \ttfamily
  arXiv:hep-ph/0211375}}.

\bibitem{Fogli:2004gy}
G.~L. Fogli, E.~Lisi, A.~Mirizzi, and D.~Montanino, ``{\em {Three generation
  flavor transitions and decays of supernova relic neutrinos}},''
  \href{http://dx.doi.org/10.1103/PhysRevD.70.013001}{Phys. Rev. {\normalfont
  \bfseries D70} (2004)  013001},
\href{http://arxiv.org/abs/hep-ph/0401227}{{\normalfont \ttfamily
  arXiv:hep-ph/0401227}}.

\bibitem{Das:2011yh}
C.~R. Das and J.~Pulido, ``{\em {Neutrino nonstandard interactions in the
  supernova}},'' \href{http://dx.doi.org/10.1103/PhysRevD.84.105040}{Phys. Rev.
  {\normalfont \bfseries D84} (2011)  105040},
\href{http://arxiv.org/abs/1106.4268}{{\normalfont \ttfamily arXiv:1106.4268}}.

\bibitem{Heurtier:2016otg}
L.~Heurtier and Y.~Zhang, ``{\em {Supernova Constraints on Massive
  (Pseudo)Scalar Coupling to Neutrinos}},''
  \href{http://dx.doi.org/10.1088/1475-7516/2017/02/042}{JCAP {\normalfont
  \bfseries 1702} (2017)  042},
\href{http://arxiv.org/abs/1609.05882}{{\normalfont \ttfamily
  arXiv:1609.05882}}.

\bibitem{Farzan:2018gtr}
Y.~Farzan, M.~Lindner, W.~Rodejohann, and X.-J. Xu, ``{\em {Probing neutrino
  coupling to a light scalar with coherent neutrino scattering}},''
  \href{http://dx.doi.org/10.1007/JHEP05(2018)066}{JHEP {\normalfont \bfseries
  05} (2018)  066},
\href{http://arxiv.org/abs/1802.05171}{{\normalfont \ttfamily
  arXiv:1802.05171}}.

\bibitem{Chang:2016ntp}
J.~H. Chang, R.~Essig, and S.~D. McDermott, ``{\em {Revisiting Supernova 1987A
  Constraints on Dark Photons}},''
  \href{http://dx.doi.org/10.1007/JHEP01(2017)107}{JHEP {\normalfont \bfseries
  01} (2017)  107},
\href{http://arxiv.org/abs/1611.03864}{{\normalfont \ttfamily
  arXiv:1611.03864}}.

\bibitem{Chang:2018rso}
J.~H. Chang, R.~Essig, and S.~D. McDermott, ``{\em {Supernova 1987A Constraints
  on Sub-GeV Dark Sectors, Millicharged Particles, the QCD Axion, and an
  Axion-like Particle}},''
  \href{http://dx.doi.org/10.1007/JHEP09(2018)051}{JHEP {\normalfont \bfseries
  09} (2018)  051},
\href{http://arxiv.org/abs/1803.00993}{{\normalfont \ttfamily
  arXiv:1803.00993}}.

\bibitem{Mangano:2011ar}
G.~Mangano and P.~D. Serpico, ``{\em {A robust upper limit on $N_{\rm eff}$
  from BBN, circa 2011}},''
  \href{http://dx.doi.org/10.1016/j.physletb.2011.05.075}{Phys. Lett.
  {\normalfont \bfseries B701} (2011)  296--299},
\href{http://arxiv.org/abs/1103.1261}{{\normalfont \ttfamily arXiv:1103.1261}}.

\bibitem{Ahlgren:2013wba}
B.~Ahlgren, T.~Ohlsson, and S.~Zhou, ``{\em {Comment on “Is Dark Matter with
  Long-Range Interactions a Solution to All Small-Scale Problems of $\Lambda$
  Cold Dark Matter Cosmology?”}},''
  \href{http://dx.doi.org/10.1103/PhysRevLett.111.199001}{Phys. Rev. Lett.
  {\normalfont \bfseries 111} (2013)  199001},
\href{http://arxiv.org/abs/1309.0991}{{\normalfont \ttfamily arXiv:1309.0991}}.

\bibitem{Huang:2017egl}
G.-y. Huang, T.~Ohlsson, and S.~Zhou, ``{\em {Observational Constraints on
  Secret Neutrino Interactions from Big Bang Nucleosynthesis}},''
  \href{http://dx.doi.org/10.1103/PhysRevD.97.075009}{Phys. Rev. {\normalfont
  \bfseries D97} (2018)  075009},
\href{http://arxiv.org/abs/1712.04792}{{\normalfont \ttfamily
  arXiv:1712.04792}}.

\bibitem{Aghanim:2018eyx}
{\normalfont \bfseries Planck}, N.~Aghanim {\em et al.}, ``{\em {Planck 2018
  results. VI. Cosmological parameters}},''
\href{http://arxiv.org/abs/1807.06209}{{\normalfont \ttfamily
  arXiv:1807.06209}}.

\bibitem{Boehm:2012gr}
C.~Boehm, M.~J. Dolan, and C.~McCabe, ``{\em {Increasing $N_{\rm eff}$ with
  particles in thermal equilibrium with neutrinos}},''
  \href{http://dx.doi.org/10.1088/1475-7516/2012/12/027}{JCAP {\normalfont
  \bfseries 1212} (2012)  027},
\href{http://arxiv.org/abs/1207.0497}{{\normalfont \ttfamily arXiv:1207.0497}}.

\bibitem{Hannestad:2004qu}
S.~Hannestad, ``{\em {Structure formation with strongly interacting neutrinos -
  Implications for the cosmological neutrino mass bound}},''
  \href{http://dx.doi.org/10.1088/1475-7516/2005/02/011}{JCAP {\normalfont
  \bfseries 0502} (2005)  011},
\href{http://arxiv.org/abs/astro-ph/0411475}{{\normalfont \ttfamily
  arXiv:astro-ph/0411475}}.

\bibitem{Bell:2005dr}
N.~F. Bell, E.~Pierpaoli, and K.~Sigurdson, ``{\em {Cosmological signatures of
  interacting neutrinos}},''
  \href{http://dx.doi.org/10.1103/PhysRevD.73.063523}{Phys. Rev. {\normalfont
  \bfseries D73} (2006)  063523},
\href{http://arxiv.org/abs/astro-ph/0511410}{{\normalfont \ttfamily
  arXiv:astro-ph/0511410}}.

\bibitem{Cyr-Racine:2013jua}
F.-Y. Cyr-Racine and K.~Sigurdson, ``{\em {Limits on Neutrino-Neutrino
  Scattering in the Early Universe}},''
  \href{http://dx.doi.org/10.1103/PhysRevD.90.123533}{Phys. Rev. {\normalfont
  \bfseries D90} (2014)  123533},
\href{http://arxiv.org/abs/1306.1536}{{\normalfont \ttfamily arXiv:1306.1536}}.

\bibitem{Archidiacono:2013dua}
M.~Archidiacono and S.~Hannestad, ``{\em {Updated constraints on non-standard
  neutrino interactions from Planck}},''
  \href{http://dx.doi.org/10.1088/1475-7516/2014/07/046}{JCAP {\normalfont
  \bfseries 1407} (2014)  046},
\href{http://arxiv.org/abs/1311.3873}{{\normalfont \ttfamily arXiv:1311.3873}}.

\bibitem{Lancaster:2017ksf}
L.~Lancaster, F.-Y. Cyr-Racine, L.~Knox, and Z.~Pan, ``{\em {A tale of two
  modes: Neutrino free-streaming in the early universe}},''
  \href{http://dx.doi.org/10.1088/1475-7516/2017/07/033}{JCAP {\normalfont
  \bfseries 1707} (2017)  033},
\href{http://arxiv.org/abs/1704.06657}{{\normalfont \ttfamily
  arXiv:1704.06657}}.

\bibitem{Oldengott:2017fhy}
I.~M. Oldengott, T.~Tram, C.~Rampf, and Y.~Y.~Y. Wong, ``{\em {Interacting
  neutrinos in cosmology: exact description and constraints}},''
  \href{http://dx.doi.org/10.1088/1475-7516/2017/11/027}{JCAP {\normalfont
  \bfseries 1711} (2017)  027},
\href{http://arxiv.org/abs/1706.02123}{{\normalfont \ttfamily
  arXiv:1706.02123}}.

\bibitem{Forastieri:2015paa}
F.~Forastieri, M.~Lattanzi, and P.~Natoli, ``{\em {Constraints on secret
  neutrino interactions after Planck}},''
  \href{http://dx.doi.org/10.1088/1475-7516/2015/07/014}{JCAP {\normalfont
  \bfseries 1507} (2015)  014},
\href{http://arxiv.org/abs/1504.04999}{{\normalfont \ttfamily
  arXiv:1504.04999}}.

\bibitem{Nyborg:1965zz}
P.~Nyborg, H.~S. Song, W.~Kernan, and R.~H. Good, ``{\em {Phase-Space
  Considerations for Four-Particle Final States}},''
\href{http://dx.doi.org/10.1103/PhysRev.140.B914}{Phys. Rev. {\normalfont
  \bfseries 140} (1965)  B914--B920}.

\end{thebibliography}\endgroup

\end{document}